\documentclass[11pt]{article}
\pdfoutput=1  
\usepackage{jcappub}
\usepackage{graphicx}
\usepackage{amsmath,amssymb}
\graphicspath{{./figures/}}
\usepackage{appendix}
\usepackage{placeins}
\usepackage[normalem]{ulem}
\usepackage[utf8]{inputenc}
\usepackage{amsfonts}
\usepackage{upgreek}
\usepackage{latexsym}
\usepackage{stfloats}
\usepackage{afterpage}
\usepackage{caption}
\usepackage{subcaption}
\usepackage{url}

\newcommand{\be}{\begin{equation}}
\newcommand{\ee}{\end{equation}}
\newcommand{\beq}{\begin{equation}}
\newcommand{\eeq}{\end{equation}}
\newcommand{\bea}{\begin{eqnarray}}
\newcommand{\eea}{\end{eqnarray}}

\newcommand{\bn}{{\mathbf  n}}
\newcommand{\bk}{{\mathbf k}}

\newcommand{\HH}{{\cal H}}

\newcommand{\De}{\Delta}

\newcommand{\ka}{\kappa}

\newcommand{\la}{\lambda}

\newcommand{\si}{\sigma}

\definecolor{magenta}{rgb}{0.1,0.98,0.6}

\definecolor{dgreen}{rgb}{0,0.7,0.0}

\title{Cosmological Simulations of Number Counts}
\author{Francesca Lepori${}^\text{1,2}$,}
\author{Julian Adamek${}^\text{2}$}
\author{and Ruth Durrer${}^\text{1}$}
\affiliation{${}^\text{1}$Universit\'e de Gen\`eve, D\'epartement de Physique Th\'eorique and Centre for Astroparticle Physics, 24 quai Ernest-Ansermet, CH-1211 Gen\`eve 4, Switzerland}
\affiliation{${}^\text{2}$Universit\"at Z\"urich, Institute for Computational Science, Winterthurerstrasse 190, CH-8057 Z\"urich, Switzerland}

\emailAdd{francesca.lepori2@uzh.ch}
\emailAdd{julian.adamek@uzh.ch}
\emailAdd{ruth.durrer@unige.ch}

\abstract{In this paper we present for the first time the angular power spectra $C_\ell(z,z')$ for number counts from relativistic 
N-body simulations. We use the relativistic N-body code {\em gevolution} with its exact integration of lightlike geodesics which include all relativistic scalar contributions to the number counts. We compare our non-perturbative numerical results with the results from {\sc class} using the {\sc hmcode} approximation for the non-linear matter power spectrum. We find that this simple description is excellent for both, the density and the convergence.
On the other hand, the current implementation of redshift-space distortions in Boltzmann codes is not accurate.
We also find that the largest contribution to the unequal-redshift power spectra is the cross-correlation of the density and the lensing contribution to the number counts, especially for redshift bins that are far apart. Correlating 
the number counts with the convergence map we find that the signal is dominated by the lensing-lensing term  when the convergence field redshift is not higher than the number counts one, while it is dominated by the density-lensing term in the opposite case.
In the present study, the issue of galaxy bias is deliberately left aside by considering only unbiased samples of matter particles from the simulations.
}

\begin{document}

\maketitle

\section{Introduction}
In the coming years several large and deep surveys of galaxies will be carried out. Most notably the photometric survey DES which has completed five years of operation~\cite{Abbott:2017wau,Abbott:2021bzy}, the spectroscopic survey DESI~\cite{Aghamousa:2016zmz} which is already taking data, the Euclid satellite survey~\cite{Laureijs:2011gra,Amendola:2016saw} slated for launch in 2022, the Vera C. Rubin Observatory's Legacy Survey of Space and Time (LSST) which will start taking data in about two years~\cite{Abell:2009aa,Abate:2012za}, the Square Kilometer Array (SKA)~\cite{Maartens:2015mra,Santos:2015gra} which is under construction in South Africa and Australia, and SPHEREX~\cite{Dore:2014cca}. These surveys will contain many millions of galaxies out to redshift $z\simeq 3$ and it is of great importance that we profit optimally from this treasure of data to learn about the parameters describing our Universe, but also to test General Relativity itself on cosmological scales.

For this reason an accurate modelling of the distribution of galaxies and their correlation properties is needed. Since we shall be able to measure the correlation function out to very large scales, we want to include relativistic projection effects which stem from the fact that measurements are made on our past light cone which is itself perturbed by the clustering of matter. Furthermore, the positions of galaxies are measured as redshifts and angles which are affected by the proper motion of galaxies and by lensing. These effects have been taken into account fully within linear perturbation theory calculations, see Refs.~\cite{2009PhRvD..80h3514Y,Bonvin:2011bg,Challinor:2011bk,DiDio:2013bqa} for a complete analysis and~\cite{Kaiser1987,Matsubara:2004fr} for relevant partial results.  However, at late times and on intermediate to small scales, density perturbations grow large and linear perturbation theory is not sufficient. Some attempts to use second order relativistic perturbation theory have been made, but this seems to be quite cumbersome~\cite{Yoo:2014sfa,Umeh:2014ana,DiDio:2014lka}, see also~\cite{Nielsen:2016ldx} for a significant simplification.

In numerical simulations, large-scale structure observables have been studied in the non-linear regime
with Newtonian simulations, see e.g.\ \cite{Fosalba:2007mf, Hilbert:2008kb, Fosalba:2013wxa, Fosalba:2013mra}.
However, in the past years there has been a growing interest in including relativistic effects in this framework \cite{Borzyszkowski:2017ayl, Zhu:2017jfl, Giblin:2017ezj, Breton:2018wzk, Beutler:2020evf, Lepori:2020ifz, Guandalin:2020snp, Coates:2020jzw}. 
Some of those have been detected using Newtonian N-body simulations, e.g. by
measuring the dipole of the correlation function and of the power spectrum when two halo populations are cross-correlated.

In this paper, we investigate the non-linear regime of relativistic effects on the angular power spectrum of the number counts, extracted directly from relativistic N-body simulations using the code \textit{gevolution}~\cite{Adamek:2015eda,Adamek:2016zes}. 

 In this code, the metric perturbations themselves are included at first order, while their first and second spatial derivatives are included to all orders. Recently, this code has been used to calculate the lensing shear, magnification and rotation by integrating photon geodesics~\cite{Lepori:2020ifz}. Integrating the geodesics exactly to all orders in scalar metric perturbations, we produce mock catalogues giving the observed direction and redshift of chosen point sources. 

Since our simulations do not capture the complicated astrophysical process of galaxy formation we sample our point sources directly from the N-body ensemble, providing an unbiased tracer of matter.
Compared to previous numerical studies in the literature,
the ray-tracing method that we use is non-perturbative, i.e.\ it does not rely on the Born approximation and it includes all relativistic scalar contributions.

The paper is organised as follows. In the next section we briefly introduce the concept of number counts from the point of view of cosmological perturbation theory. In Section~\ref{s:sim} we describe the simulations performed for this work as well as the numerical processing of the data. In Section~\ref{s:res} we present and discuss our results and in Section~\ref{s:con} we conclude. Technical details about the impact of the survey window are presented in an appendix.

We consider a  flat Friedmann-Lema\^itre-Robertson-Walker (FLRW) universe where the metric is written in Poisson gauge as
\begin{equation}
ds^2 = a^2(\tau)\! \left[-e^{2\psi} d\tau^2 \!-\! 2 B_i dx^i d\tau \!+\! \left(e^{-2\phi} \delta_{ij} \!+\! h_{ij}\right)\!dx^i dx^j\right]\,,
\end{equation}
where $\tau$ is the conformal time, $a(\tau)$ is the background scale factor and 
$x^i$ are comoving Cartesian coordinates. Here, $\psi$ and $\phi$ are the two scalar gravitational
potentials, $B_i$ is the divergenceless vector potential that is responsible for frame dragging, and $h_{ij}$ is a transverse and traceless spin-2 field that denotes the tensor degrees of freedom of the metric.

\section{Linear perturbation theory}
Large-scale structure observations map the matter distribution across cosmic history measuring the fluctuations of tracers across the sky at different redshifts. For discrete tracers, e.g.\ galaxies or quasars, the observable 
from which we extract this information
is the \emph{number counts}, the fractional overdensity of the number of observed objects in
a given redshift bin,
\begin{equation}
    \Delta(z, \mathbf{n}) = \frac{N(z, \mathbf{n}) - \bar{N}(z)}{\bar{N}(z)},
\end{equation}
where $\mathbf{n}$ is the direction of observation,
$z$ is the observed redshift, while $N(z, \mathbf{n})$
and $\bar{N}(z)$ denote the number of objects per solid angle seen at position $\mathbf{n}$ and the average number of objects detected over the whole survey sky, respectively. 
Since sources lie on the past light cone of the observer, a fully relativistic prediction for the number counts 
not only encodes information on the
underlying dark matter fluctuations, but also on the tracers' peculiar velocities 
and the geometry of the spacetime itself through lensing, integrated Sachs-Wolfe and time-delay
effects. In linear perturbation theory the observed number counts can be schematically expressed as a sum of different contributions \cite{2009PhRvD..80h3514Y,Bonvin:2011bg,Challinor:2011bk,DiDio:2013bqa}
\begin{equation}
\label{1}
\Delta= \Delta_\text{dens} + \Delta_\text{rsd} + \Delta_{\kappa} + \Delta_\text{gr},
\end{equation}
where $\Delta_\text{dens}$ is the local overdensity,
\begin{equation}
  \Delta_\text{dens} = b  \  \delta_\text{cm}
  \label{eq:dens}
\end{equation}
proportional to the density contrast in comoving gauge  $\delta_\text{cm}$. 
For unbiased tracers, the bias is simply $b = 1$. 
In our notation, the term $\Delta_\text{rsd}$ includes the linear Kaiser redshift-space distortions (RSD)~\cite{Kaiser1987}, plus other subdominant Doppler corrections
\begin{equation}
\Delta_\text{rsd} = \frac{1}{\mathcal{H}(z)} \partial_r (\mathbf{v}\cdot\mathbf{n}) + \Biggl(\frac{\mathcal{H'}}{\mathcal{H}^2} + \frac{2}{r\mathcal{H}}\Biggr)(\mathbf{v}\cdot \mathbf{n})
 - 3\HH V , \label{RSDcorr}
\end{equation}
where $\mathbf{v}$ is the peculiar velocity in Poisson gauge, $V$ the velocity potential defined by $\mathbf{v} = -{\bf \nabla} V$, $r = \tau_o -\tau$ is the conformal distance,
$\mathcal{H}=a'/a$ is the conformal Hubble factor and a prime denotes a derivative with respect to the conformal time $\tau$.
$\Delta_{\kappa}$ is the gravitational lensing term,
\begin{equation}
\Delta_{\kappa} = -\int_0^{r(z)} \frac{r(z) - r}{r(z) r} \Delta_{\Omega} (\Phi + \Psi) dr,
\label{eq:kappa}
\end{equation}
where $\Delta_{\Omega}$
is the Laplace operator on the sphere.
The last term in Eq.~\eqref{1}, $ \Delta_{\rm gr},$ includes
local and integrated combinations of the Bardeen potentials:
\begin{align}
\Delta_{\rm gr} = &- 2 \Phi + \Psi + \frac{1}{\mathcal{H}} \Phi'
              +\frac{2}{r(z)} \int^{r(z)}_0 dr (\Phi + \Psi) + \nonumber  \\
              & +\Biggl(\frac{\mathcal{H'}}{\mathcal{H}^2} + \frac{2}{r(z)\mathcal{H}}\Biggr)\Biggl(\Psi + \int^{r(z)}_0 dr (\Phi' + \Psi')\Biggr)\, . \label{Delta_rel}
\end{align}

The density and the Kaiser redshift-space distortions are often referred to as \emph{standard} terms, since current clustering analysis take them into account in the physical model for the observable.  
The gravitational lensing contribution was also
well known before the full computation of the relativistic number counts was carried out and it has been detected for the first time in the Sloan Digital Sky Survey (SDSS) by cross-correlating quasars and their foreground galaxies \cite{Scranton:2005ci}. More recently, Ref.~\cite{Liu:2021gbm} presented a detection of the lensing signal from the cross-correlation of the background galaxies and their foreground convergence field. This effect has been investigated in detail in the literature and the general consensus is that it will be crucial to model it in the analysis of future cosmological surveys \cite{Duncan:2013haa, Montanari:2015rga, Raccanelli:2015vla, Cardona:2016qxn, DiDio:2016ykq, Lorenz:2017iez, Villa_2018, Tanidis:2019fdh, Jelic-Cizmek:2020pkh, Bellomo:2020pnw}. 
The Doppler contribution to Eq.~\eqref{RSDcorr}
is sub-dominant in the standard clustering analysis. However, it is possible to isolate this term from the density and Kaiser contributions measuring the odd multipoles of the Fourier space power spectrum and correlation function using multiple tracer techniques~\cite{McDonald_2009, Bonvin:2013ogt}. This has been investigated with N-body simulation in Refs.~\cite{Breton:2018wzk, Beutler:2020evf, Guandalin:2020snp} and detected from galaxy clustering measurements in Ref.~\cite{Alam:2017izi}. 
Finally, the perspective of detection for the terms grouped in Eq.~\eqref{Delta_rel} has also been investigated in the literature and the present forecasts suggest that 
cosmological probes will not be able to see them in the foreseeable future, at least not in a single-tracer analysis \cite{Alonso:2015uua}.

We remark that the full number counts
are a gauge-invariant quantity. However, 
the contributions that we single out 
here in order to separate the different physical effects that come into play
are not all gauge invariant separately and, therefore, cannot be observed individually.
The lensing contribution $\Delta_\kappa$ can be measured independently by weak-lensing observations and is indeed gauge invariant, however, we note that additional lensing-like effects can occur in different gauges, see \cite{Adamek:2019aad} for an example.
Throughout this work we will consider 
the ensembles of N-body particles in our simulations as tracers of the 
distribution of dark matter and baryons. Therefore, we can 
safely neglect the effects of clustering bias, magnification bias and evolution bias.

\section{Numerical Methods}\label{s:sim}

Our numerical study uses four different mock catalogues from three separate N-body simulations that all use exactly the same baseline cosmology. This allows us to cover various choices of resolution, simulation volume, survey area and redshift range so that we can check the robustness of our results with respect to such choices. One simulation has a relatively large volume of $(4032~\mathrm{Mpc}/h)^3$ and therefore correspondingly a low resolution of just $700~\mathrm{kpc}/h$ (we used $5760^3$ grid points and the same number of particles) which yields a mass resolution of $3 \times 10^{10}~M_\odot/h$. This simulation, called \texttt{unity-2}, provides two mock surveys for the same observer location: a full-sky survey out to redshift $z \sim 0.85$ and a wide ``pencil-beam'' survey covering $1932$ square degrees with a redshift range from $z \sim 0.85$ to $z \sim 3.5$. In our second simulation, called \texttt{high-resolution}, we used a somewhat smaller volume of $(1920~\mathrm{Mpc}/h)^3$ but significantly higher resolution of $250~\mathrm{kpc}/h$ ($7680^3$ grid points and particles) which yields a mass resolution better than $1.4\times 10^9~M_\odot/h$. This simulation provides us with a mock survey of $1009$ square degrees out to redshift $z \sim 2$. The large redshift range is achieved by allowing the pencil-beam footprint of the survey to ``wrap around'' the periodic simulation box, however, making sure that no part of the simulation volume appears on the footprint more than once. This can be achieved by carefully choosing the field of view in the periodic setting. Our third simulation, called \texttt{high-resolution-2}, uses exactly the same resolution as the previous one, but has a somewhat smaller volume of $(1440~\mathrm{Mpc}/h)^3$ (hence using only $5760^3$ grid points and particles). It provides us with another mock survey of $1009$ square degrees that is statistically independent of the previous one. However, due to the smaller box size, it reaches a maximum redshift of only $z \sim 1.25$. The properties of our mock surveys are summarised in Table~\ref{tab:catalogues}.

\begin{table}
\begin{center}
\begin{tabular}{|c| c c c c c| } 
 \hline
 Catalogue & $z$ range & $f_\text{sky}$ & $N_\text{sources}$ & grid resolution & $L_\text{box}$\\ 
 \hline
\texttt{unity-2-low-z}   & $[0, 0.85]$ & 1 & $1.55 \times 10^9$ & $700~\mathrm{kpc}/h$ & $4032~\mathrm{Mpc}/h$\\ 
\texttt{unity-2-high-z}  & $[0.85, 3.5]$ & 0.047 & $835 \times 10^6$& $700~\mathrm{kpc}/h$ & $4032~\mathrm{Mpc}/h$\\ 
\texttt{high-resolution}  & $[0, 2]$ & 0.025 & $594 \times 10^6$& $250~\mathrm{kpc}/h$ & $1920~\mathrm{Mpc}/h$\\ 
\texttt{high-resolution-2}  & $[0, 1.25]$ & 0.025 & $251 \times 10^6$& $250~\mathrm{kpc}/h$ & $1440~\mathrm{Mpc}/h$ \\ 
 \hline
\end{tabular}
\caption{Summary of the properties of the catalogues used in our analysis: redshift range ($z$ range), sky fraction ($f_\text{sky}$), number of sources ($N_\text{sources}$), grid resolution and size of the simulation box ($L_\text{box}$).} 
\label{tab:catalogues}
\end{center}
\end{table}

The assumed baseline cosmology is a $\Lambda$CDM model with three neutrino species in a minimal-mass configuration (one massless state, $m_1 = 0~\mathrm{eV}$, and two massive states in normal hierarchy, $m_2 = 0.008689~\mathrm{eV}$ and $m_3 = 0.05~\mathrm{eV}$). The neutrinos are treated as a linear component to the stress-energy similar to how it is done in \cite{Adamek:2017grt} for the massless case (based on an idea proposed in \cite{Brandbyge:2008js}). For neutrinos with small masses this is an excellent approximation. Our N-body ensemble therefore captures the distribution of cold dark matter and baryons while total matter would include a small correction due to the massive neutrinos. The remaining cosmological parameters are $h = 0.67$, $\Omega_c = 0.26858$, $\Omega_b = 0.049$, $T_\mathrm{CMB} = 2.7255~\mathrm{K}$, and we use a nearly scale-invariant spectrum of primordial perturbations given by $A_s = 2.215\times 10^{-9}$ and $n_s = 0.9619$ at the pivot scale of $0.05~\mathrm{Mpc}^{-1}$.

Our simulations are performed with the relativistic cosmological N-body code \textit{gevolution} which provides metric perturbations and particle phase-space coordinates in Poisson gauge. This is naturally a weak-field gauge in cosmology which facilitates an efficient numerical integration of Einstein's equations that is at the same time very accurate. In the weak-field expansion employed in \textit{gevolution} all the metric variables are kept at least to first order. In addition, terms that contain spatial gradients of the scalar gravitational potentials are kept to all orders, see \cite{Adamek:2015eda,Adamek:2016zes} for more details. However, this is only necessary to ensure a consistent calculation of relativistic corrections like frame dragging or gravitational slip which are very small effects in the cosmological model considered here. In particular, the effect of frame dragging (the largest correction) on the observables we want to compute remains negligible (well below $1 \%$) and shall therefore be ignored here. Thus, for the purpose of this work, the main advantage of using the computational framework of \textit{gevolution} is the fact that the simulation output has an unambiguous relativistic interpretation and can be processed efficiently using a validated pipeline.

\begin{figure}
\begin{center}
\includegraphics[width=0.8\textwidth]{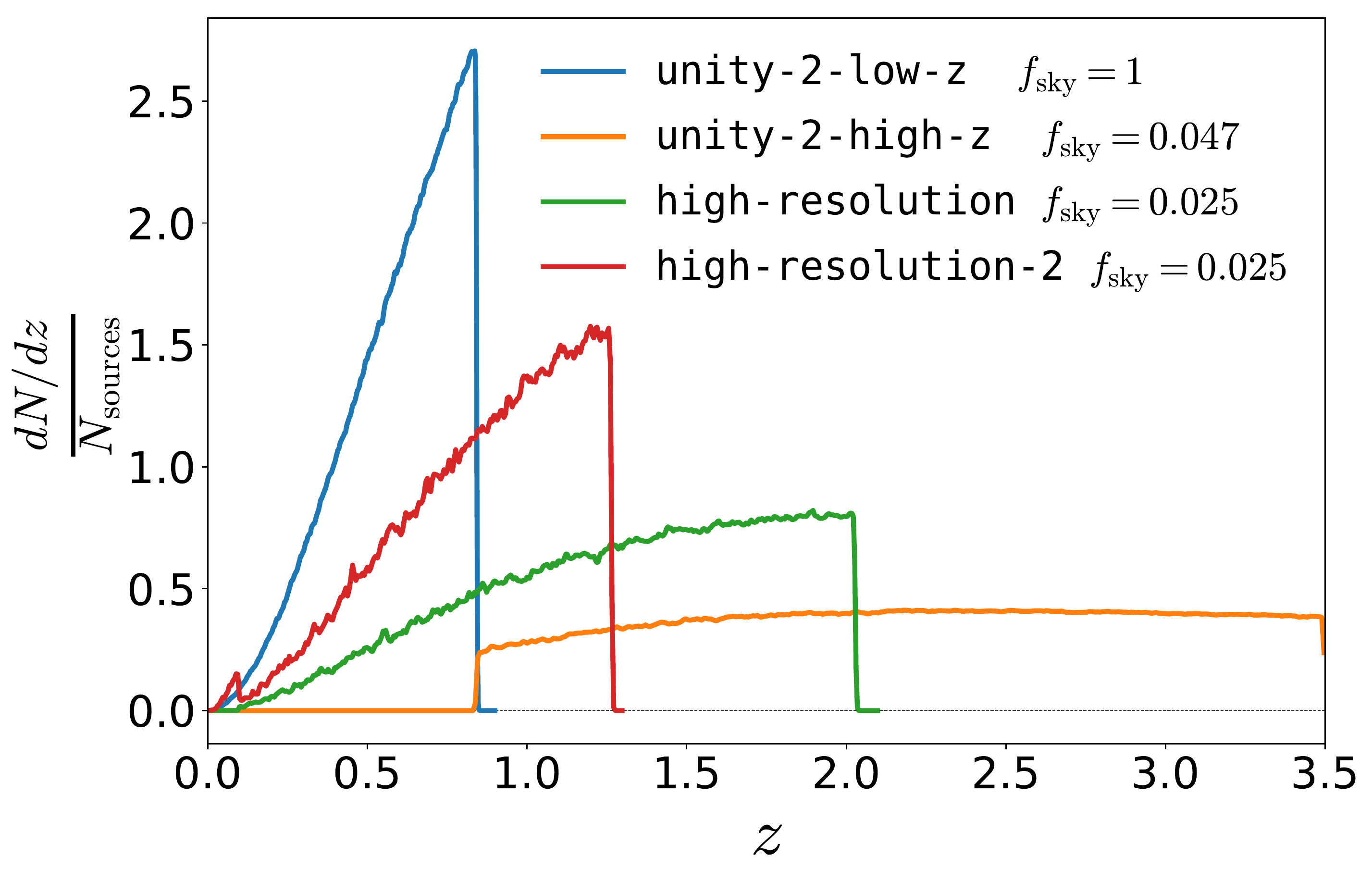}   
    \caption{The redshift distribution of the four point-source catalogues analysed in this work. The integral of the distributions is normalised to 1 in this plot.
\label{f:dNdz}    }
\end{center}
\end{figure}

From the particle and metric data on the past light cone we can construct the null geodesics that connect a catalogue of point sources with the observation event in each simulation. This provides us with observed positions and observed redshifts that include all relativistic and projection effects nonperturbatively. More details on the ray-tracing method employed here can be found in \cite{Lepori:2020ifz}. Our point-source catalogues consist of randomly drawn (and hence unbiased) subsamples of the N-body particles that represent cold dark matter and baryons. The reason for this choice is twofold. First, it allows us to leave aside the issue of bias which could otherwise complicate our analysis significantly. Second, we can work with a high density of point sources to beat down shot noise even at high redshift. The number of point sources, $N_\mathrm{source}$, in each catalogue is given in Table~\ref{tab:catalogues}, and their distribution in redshift is illustrated in Figure \ref{f:dNdz}. 

Once the angular positions and observed redshifts of all point sources have been determined we proceed with an estimation of angular power spectra $C_\ell(z, z')$ of the observed number counts as follows:
\begin{enumerate}
    \item We select all the particles within a chosen range of observed redshifts (a redshift bin with ``top-hat'' selection function) and we 
project the number of particles $n_\text{part}$
onto a sky map pixelised with the \texttt{HEALPix} method \cite{Gorski:2004by}. A Boolean mask is applied according to the survey geometry for pencil-beam surveys.
    \item We compute the average of the number of particles per pixel $\bar{n}$. Note that this average is performed only over the unmasked pixels.
    \item We compute the number-count fluctuation in each pixel $i$  as
$\Delta^i \equiv (n_\text{part}^i - \bar{n})/\bar{n}$. 
    \item We estimate the angular power spectrum for the number-count map using the code \texttt{PolSpice} \cite{Szapudi:2000xj, Chon:2003gx}. This code implements a fast estimator for the angular auto or cross power spectra from one or two sky map(s) and it corrects for the effect of the mask(s) (see Appendix~\ref{a:win} for more details). In order to reduce the statistical noise, we do not resolve the individual multipoles. Instead we bin the angular power spectra 
    in bandpowers with $\Delta \ell \approx f_\text{sky}^{-1}$.

\end{enumerate}

For the catalogues \texttt{high-resolution} and \texttt{high-resolution-2} we additionally compute the area distance at the observed position of the sources. For these catalogues, therefore, we also estimate convergence maps from the fluctuations of the area distance, exactly as described in Ref.~\cite{Lepori:2020ifz}. The cross-correlation of number counts and convergence maps is discussed in Section \ref{sec:conv-cross}.

We shall also compare our simulated spectra with predictions from perturbation theory computed with the Cosmic Linear Anisotropy Solving System ({\sc class}) code \cite{Lesgourgues:2011re, Blas:2011rf}, which calculates 
the fully relativistic number counts in linear perturbation theory~\cite{DiDio:2013bqa}. 
Non-linearities in the matter density are included in our theoretical predictions using the {\sc hmcode} prescription implemented in {\sc class} \cite{Mead:2016zqy}, with the model parameters fitted to the Cosmic Emulator dark matter only simulation \cite{Heitmann:2013bra}.

\section{Results and discussion}\label{s:res}

\subsection{Auto-correlations in photometric bins}
In this section we investigate auto-correlations of the number counts in three redshift bins, $z = 0.5, 1.5, 3$. It is well known that the relative importance of the contributions to the number counts depends on the width of the redshift bin and on the redshift (see e.g. \cite{Bonvin:2011bg} or \cite{DiDio:2013bqa}, especially Figs.~1 and 4). However, 
we cannot choose arbitrarily narrow redshift bins: we are constrained to choose the bin width in such a way that each pixel in our map contains a sufficient number of particles in order to limit shot noise. For a comoving number density $n$,
the angular density in a bin of full width $2\si_z$ is roughly 
$$
\bar N =  n V_\text{bin}\simeq  n\frac{2\si_z}{H(z)}(1+z)^2 d_A^2(z) \,,
$$
leading to a shot-noise amplitude of $C^\mathrm{noise}_\ell = 1/\bar N \propto 1/\si_z$. 
If shot noise is larger than about 1\% of the physical $C_\ell$ we can no longer neglect it. 
 For this reason, we consider redshift bins whose half-width is $\sigma_z\in[0.01, 0.1]$, which is the typical bin size of a photometric galaxy survey.    

As a first test, we study the impact of the relativistic effects in the number counts extracted from our simulations for fixed bin size.
In order to separate the different contributions to the number counts, for each redshift bin, we compute three maps:
\begin{itemize}
    \item A map for the full number counts, where particles in each pixel are selected according to their observed redshift and observed position in the sky $(z_\text{obs}, \mathbf{n})$. The observable coordinates are computed using the ray-tracing technique described in Section~\ref{s:sim}. This map includes all the relativistic effects and it is an observable.
    \item A map which neglects the redshift perturbation contribution in the number counts, where the redshift bin is selected using the background redshift of the sources, while the pixelisation is made from the observed angular position of the sources. This map does not include number-count contributions that depend on the peculiar velocity of the sources such as RSD and Doppler effects, nor gravitational redshift perturbations, but it \textit{does} include lensing magnification.
    \item A map which only includes the density contribution to the number counts, where the comoving position of the source is translated into background redshift and angular position and used to extract the map: the particles are selected in the bin on the basis of their background redshift, and the projection in the pixelised map uses the unperturbed direction of the sources.  
\end{itemize}
Note that comoving position, peculiar velocity and lensing are each gauge dependent and hence this separation of the different effects is specific to the Poisson gauge used in this work. Only the first type of map which uses only observables is fully invariant.

The angular power spectra of auto-correlations estimated with our method are affected by shot noise.  We remove the shot noise using jackknife resampling, i.e.\ we randomly split the particles in our redshift bin into two sub-samples, we extract the number-counts map for the two sub-samples separately, and we estimate the auto-correlation by cross-correlating the two maps, see e.g.\ \cite{Inman:2015pfa}.

\begin{figure}
\captionsetup[subfigure]{labelfont={Large,bf}}
  \parbox[c]{0.2\textwidth}{\subcaption{}\label{fig:1a}}\parbox[c]{0.55\textwidth}{\includegraphics[width=0.55\textwidth]{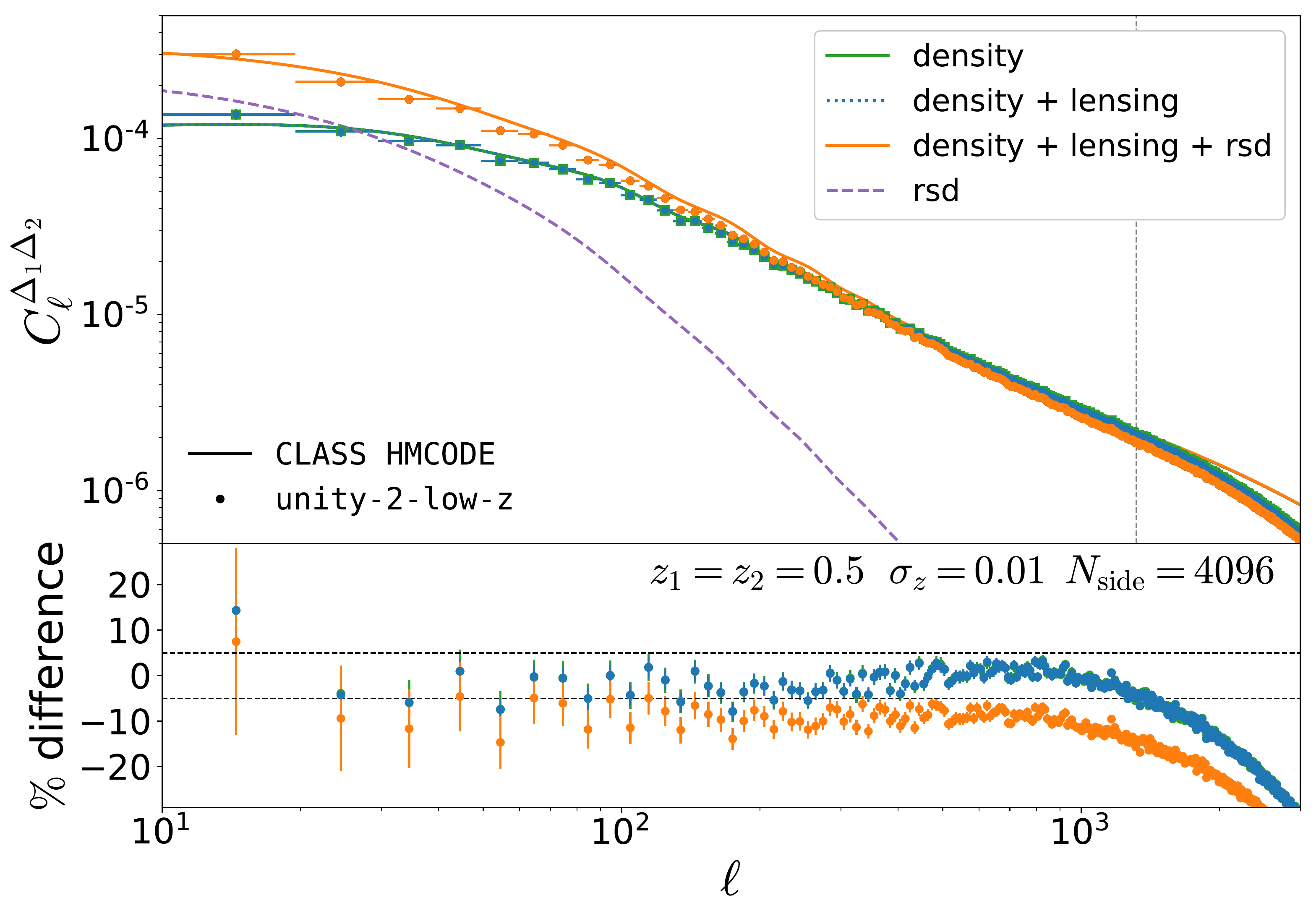}}
 \\[1ex]
    \parbox[c]{0.2\textwidth}{\subcaption{} \label{fig:1c} }\parbox[c]{0.55\textwidth}{\includegraphics[width=0.55\textwidth]{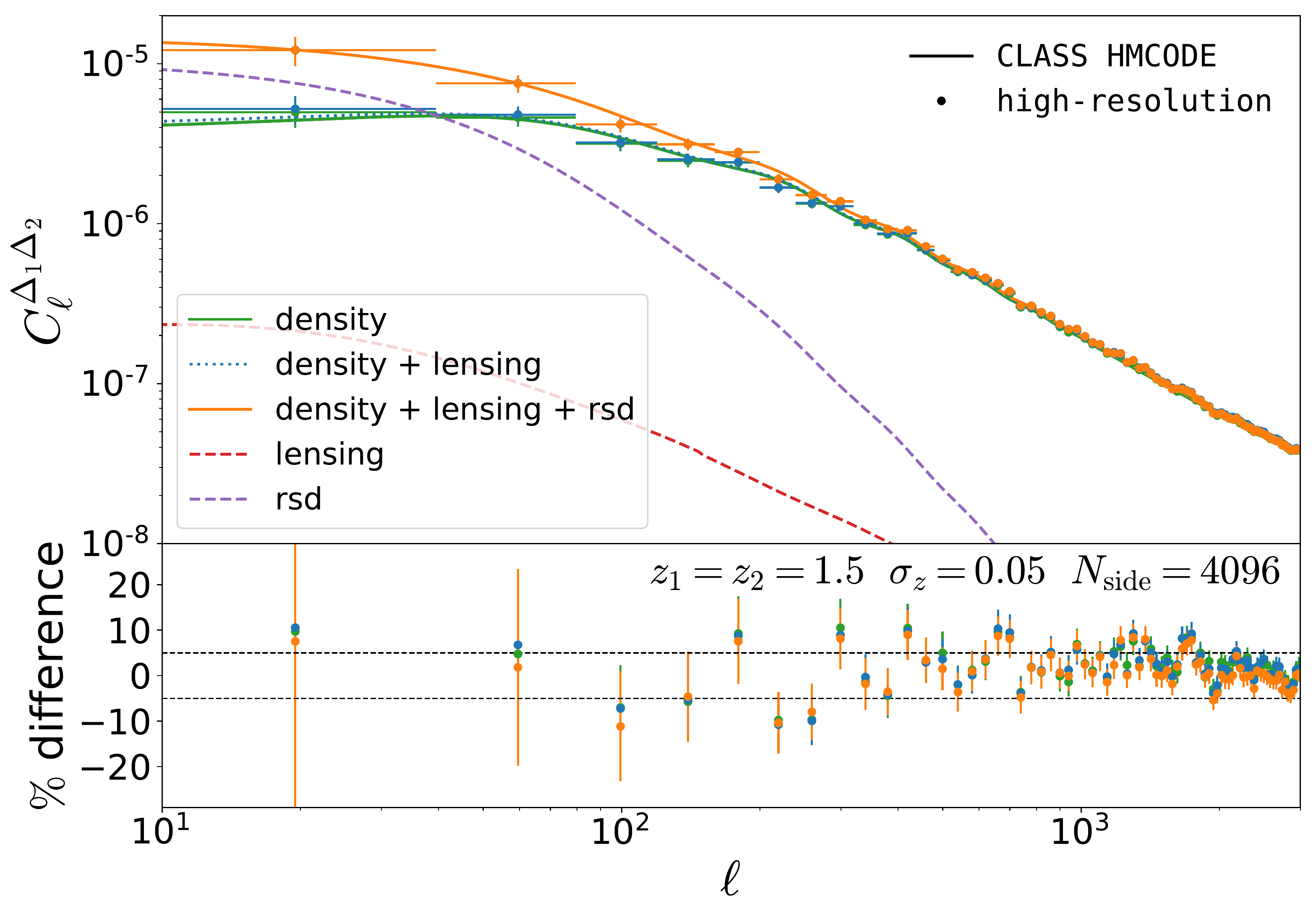}}
 \\[1ex]
   \parbox[c]{0.2\textwidth}{\subcaption{} \label{fig:1b} }\parbox[c]{0.55\textwidth}{\includegraphics[width=0.55\textwidth]{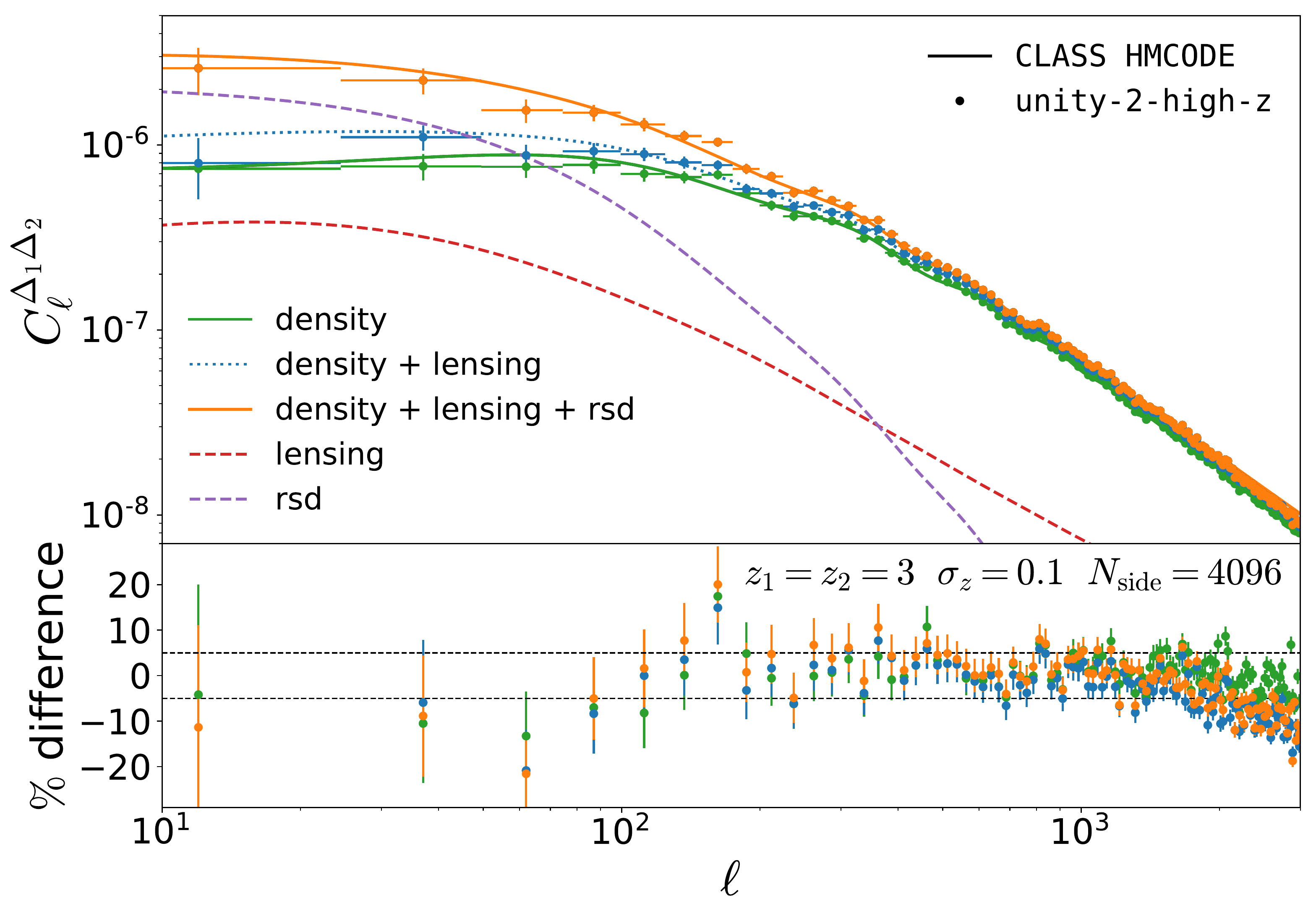}}
 \\[1ex]

        \caption{We show the equal-redshift power spectra $C_\ell(z,z)$ and the relative difference between the {\sc class hmcode}  approximation and our simulation in percent for three different redshifts and bin widths. The 5\% difference threshold is indicated by dashed lines.
        We use the \texttt{unity-2} catalogues (top and bottom panel) and the \texttt{high-resolution} catalogue (middle panel). We pixelise the sky into $12\, N_{\rm side}^2$ pixels with   $N_{\rm side}=4096$. The lensing contribution is visible by eye only for the high redshift, $z=3$.  For $z=0.5$ it is significantly smaller than 1\%, while for $z = 1.5$ the lensing contribution is below the level of  1-2\%. The \texttt{unity-2} simulation loses power at $\ell_5\simeq 1300$ for $z=0.5$, indicated as vertical line in panel (a). 
        }
        \label{fig:1}
\end{figure}
\afterpage{\FloatBarrier}

In Figs. \ref{fig:1} we compare the fully non-linear results with the theoretical prediction from {\sc class} \cite{Lesgourgues:2011re, Blas:2011rf, DiDio:2013bqa} with non-linearities modelled with {\sc hmcode}~\cite{Mead:2016zqy}. For a correct comparison with the results of our simulation, we choose top-hat selection functions with the same bin width as used to extract the number-count maps, while selection and magnification bias are set to $b = 1$ and $s = 0$, respectively, since our particles are unbiased tracers of the baryon and dark matter distribution and the catalogues are not magnitude limited.   

The three panels in Fig.~\ref{fig:1} refer to different redshift bins: $z = 0.5 \pm 0.01$ in the top panel, $z = 1.5 \pm 0.05$ in the middle panel, and $z = 3 \pm 0.1$ in the bottom panel.
Each panel shows two plots: the angular power spectra (top) and the difference (in percentage) between {\sc class} and our numerical measurement (bottom). 
Orange, blue and green markers refer, respectively, to the power spectrum extracted from the fully relativistic number-count map, the map which neglects the impact of peculiar velocities in the number counts, and the map which further neglects lensing and therefore only includes density perturbations. 

{\sc class} predictions are represented as continuous or dotted lines and their colour follows the pattern of  the simulation data: continuous orange lines include the contributions from density, lensing magnification and RSD [the three terms in Eq.~\eqref{eq:dens}, Eq.~\eqref{RSDcorr} and Eq.~\eqref{eq:kappa}], dotted blue lines includes density and lensing [the terms in Eq.~\eqref{eq:dens} and Eq.~\eqref{eq:kappa}], while continuous green lines includes only density [Eq.~\eqref{eq:dens}]. Dashed lines show the individual contribution of RSD (purple) and lensing (red), when their amplitude is large enough to be visible in the panel. Note that these contributions include also the cross-correlation of RSD/lensing with the density. We have checked that the {\sc class} result including large-scale relativistic effects ($\Delta_\mathrm{gr}$ in our notation) does not visibly differ from the orange line. 

The horizontal bars indicate our binning in multipole $\ell$, while the vertical errorbars are due to cosmic variance:
\begin{equation}
\sigma_{\ell} =
\sqrt{\frac{2}{f_\text{sky} \Delta \ell (2 \ell + 1)}} C_\ell.
\end{equation}

For all the redshift bins, we find that RSD  significantly affects the angular power spectrum on large scales ($\ell \lesssim 100$), while on small scales correlations are largely dominated by the density contribution. In the theoretical prediction, estimated with {\sc class}, we can separate the different contributions to the RSD (the Kaiser and the Doppler terms) and we find that the peculiar velocity contribution is dominated by the first one, while Doppler contributions are sub-percent even at $z=0.5$.
Lensing can be safely neglected in the auto-correlation at $z = 0.5$ and $z = 1.5$, contributing $< 0.1 \%$ and $< 2\%$, respectively. 
At high redshift ($z = 3$) the lensing contribution is of the order of $10-15\%$. 

We find that the angular power spectra extracted from simulations
agree well with the {\sc hmcode} prediction 
when RSD is not strongly affected by non-linearities. 
This is not surprising: on small scales the angular correlations are dominated by density correlations and the {\sc hmcode} is a fit for the matter power spectrum accurate at the few percent level \cite{Mead:2020vgs}.

On the other hand, as we discuss below, at low redshift non-linearities strongly affect the peculiar velocities of the particles and the contribution from RSD reduces the spectrum from simulations so that the current implementation of non-linear RSD in {\sc class} yields a somewhat too high total result on small, non-linear scales. For our example with $z=0.5$ and $\si_z=0.01$ RSD reduce the power spectrum to about 15 \% below the {\sc class} prediction around $\ell\sim 1000$.

In order to estimate the multipole $\ell$ at which resolution effects degrade the power spectrum we first estimate the effective angular resolution of our simulation for a source plane positioned at redshift $z$. This is given by $\theta_{\min} = \De x/r(z)= \pi/(k_{\max}r(z))$, where $\De x$ denotes the grid resolution and $k_{\max}=\pi/\De x$ is the corresponding Nyquist wavenumber.  This leads to a maximum multipole of
\be
\ell_{\max}(z) =\frac{\pi}{\theta_{\min}} = k_{\max}r(z)\,.
\ee
Note, however, that at this value of $\ell$ we expect errors of order unity in our simulation. A 5\% accuracy is achieved roughly at $\ell_5(z)\simeq \ell_{\max}(z)/\sqrt{20}$, since the error grows like $k^2$ when approaching the Nyquist wavenumber~\cite{colombi:2009x}. This loss of power is well visible in the density power spectrum (blue dots) in Fig.~\ref{fig:1a} beyond $\ell\simeq 2000$. 

At high redshift, finite resolution effects 
appear on smaller scales: at $z = 3$ (Fig.~\ref{fig:1}, bottom panel) grid resolution effects for the \texttt{unity-2-high-z} catalogue are $\sim 5\%$ at $\ell_5 \sim 4000$, while for the \texttt{high-resolution} simulation at $z = 1.5$ (Fig.~\ref{fig:1}, middle panel) we find
$\ell_5 \sim 8000$. Therefore, at high redshift, grid-resolution effects have negligible impact in the angular power spectra extracted from our simulations in the range of scales under consideration. We find that our simulations agree with the {\sc hmcode} prescription at the $5\%$ level up to $\ell \sim 1000$. 

\subsection{Isolating redshift-space distortions}

While density correlations can be described quite accurately 
with the {\sc class} perturbative prescription in the non-linear regime, velocity correlations are more challenging to model. 
The RSD effect is computed in {\sc class} from the Kaiser formula \cite{Kaiser1987}, $P_H(k) \mapsto P_H(k)[1+(\bn\cdot \hat\bk)^2]^2$, where $P_H$ denotes the {\sc hmcode} power spectrum, implicitly assuming the linear relation between density and velocity. However, non-linearities affect velocities and density differently. Due to angular momentum conservation velocities actually prevent infall in the non-linear regime and reduce the power spectrum with respect to the linear Kaiser prescription. Furthermore, at intermediate scales, non-linearities affect velocities more strongly than density. 
Attempts have been made to model
this reduction, e.g.\ by a Lorentzian or Gaussian kernel~\cite{Heavens:1998es,Matsubara:2007wj,Taruya:2013my} on top of one or two loop corrections from perturbation theory, or by using effective field theory~\cite{delaBella:2018fdb}. However, it has been shown that all these models of velocity non-linearities are quite inaccurate~\cite{Jalilvand:2019brk} and an accurate model for non-linear RSD, including also the small scale finger-of-god effect is still lacking. Therefore, we do not expect the current implementation in {\sc class} 
to be accurate for the velocity contribution to the angular power spectrum, especially at low redshift where non-linearities are strongest.

\begin{figure}[t!]
\begin{center}
\includegraphics[width=0.8\textwidth]{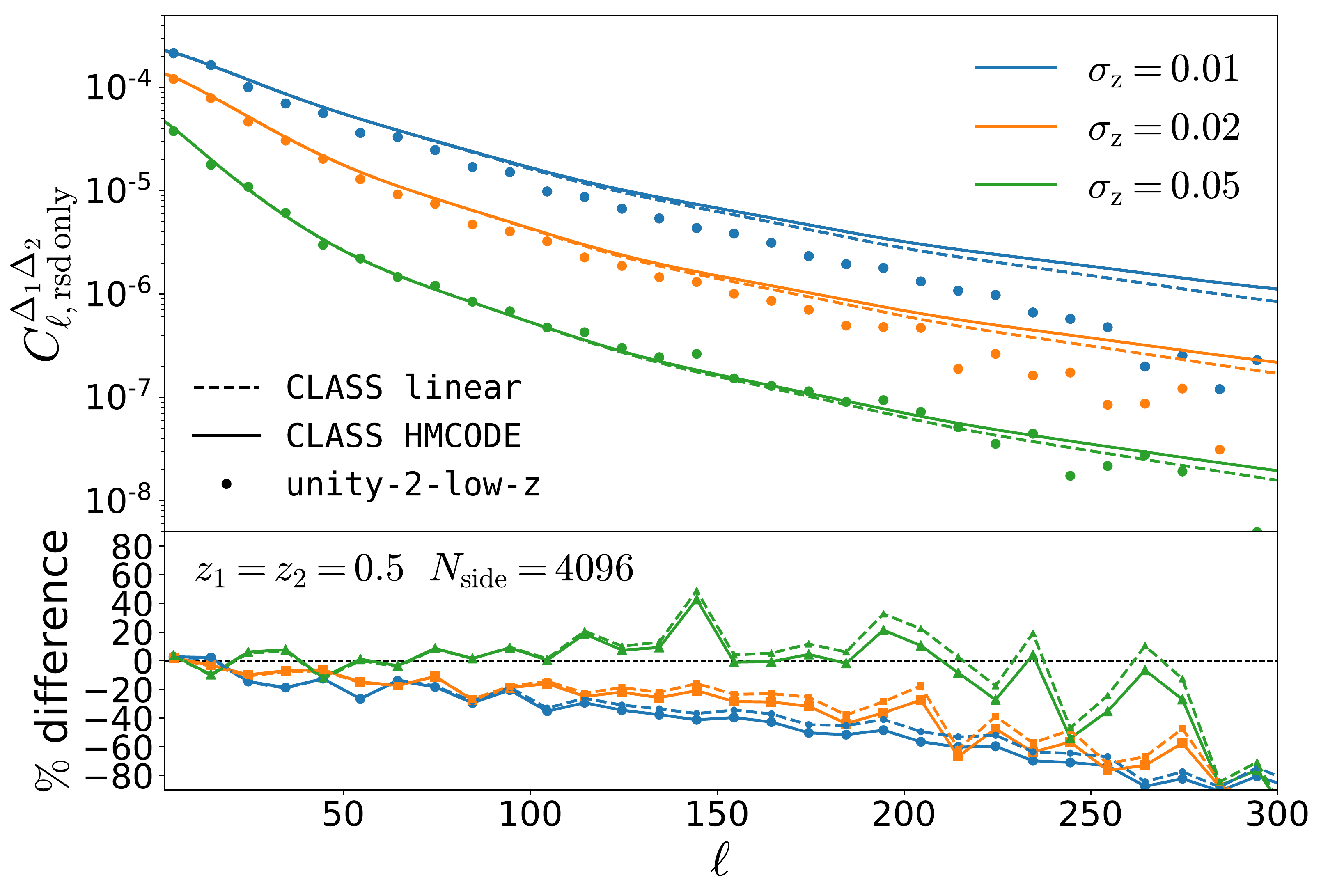}   
    \caption{The RSD power spectra at $z=0.5$ from {\sc class hmcode} (solid) and from \textit{gevolution} (dotted). In the bottom panel, we show the relative difference in \%. The {\sc class} prescription significantly overestimates the fully non-linear RSD for narrow redshift bins. For the widest redshift bin, $\si_z=0.05$, the signal is very low and noisy at $\ell>200$.\label{fig:isolate-rsd}
    }
\end{center}
\end{figure}

In order to study how non-linearities affect the RSD contribution to the angular power spectrum, we isolate RSD from the dominant density correlations by subtracting from the angular power
spectrum of the full number counts map 
the spectrum of the map that neglects redshift perturbations. 
In Fig.~\ref{fig:isolate-rsd} we compare the difference of the two spectra to the {\sc class} prediction of the same quantity, for three redshift bins centered at $z = 0.5$ with different half-widths $\sigma_\text{z} = 0.01, 0.02, 0.05$. As is well known, the narrower the redshift bin, the more pronounced is the contribution from RSD.
The figure includes scales $\ell < 300$: radial correlations are smeared over a wide bin $\De r\simeq 2\si_z/H(z)\simeq 4580\si_z h^{-1}$Mpc $>45 h^{-1}$Mpc in this analysis, and therefore the RSD contribution on smaller scales, higher $\ell$'s, is negligibly small. Furthermore, the scales considered are not affected by finite-grid resolution effects. 

On these scales, the {\sc class} prediction agrees with the results from our simulation for the largest bin-size considered here ($\sigma_\text{z} = 0.05$, $\De r\sim 230 h^{-1}$Mpc). However, for thinner
bins we find that {\sc class} overestimates the angular correlations at the level of up to $\sim 80\%$ at $\ell \gtrsim 260$. Furthermore, non-linearities of velocities are relevant at the percent level already at scales of more than $50 h^{-1}$Mpc at $z=0.5$. 
Note that in a redshift bin of half width $\si_z$ in the $C_\ell(z)$ scales from $\la_{\min}\sim \min\{\De r, {d}_A(z)/\ell\}$ to $\la_{\max}\sim \sqrt{({d}_A(z)/\ell)^2 +\De r^2(z)}$ are mixed. For $\ell=20$ and $\si_z=0.01$ at $z=0.5$ these scales are $\la\in [45,69]h^{-1}$Mpc.

These results are in good agreement with previous studies on the 
angular clustering in redshift space, based on Newtonian N-body simulations \cite{Fosalba:2013wxa}, and with the theoretical expectation that non-linear contributions to RSD are relevant already on rather large scales~\cite{Gebhardt:2020imr}.

\subsection{Extracting lensing magnification from cross-correlations}

In this section we study the unequal-redshift correlations. As is known from linear perturbation theory and is also confirmed here, the angular power spectra $C_\ell(z_1,z_2)$ for $z_1 \ne z_2$ are dominated by the density-lensing and lensing-lensing terms~\cite{Montanari:2015rga} for large redshift separation. They therefore provide an alternative route (together with  galaxy-galaxy lensing  measurements) to measure the density-lensing and lensing-lensing correlations.

\begin{figure}[ht]
\captionsetup[subfigure]{labelfont={Large,bf}}
  \parbox[c]{0.2\textwidth}{\subcaption{}\label{fig:lens-1}}\parbox[c]{0.8\textwidth}{\includegraphics[width=0.79\textwidth]{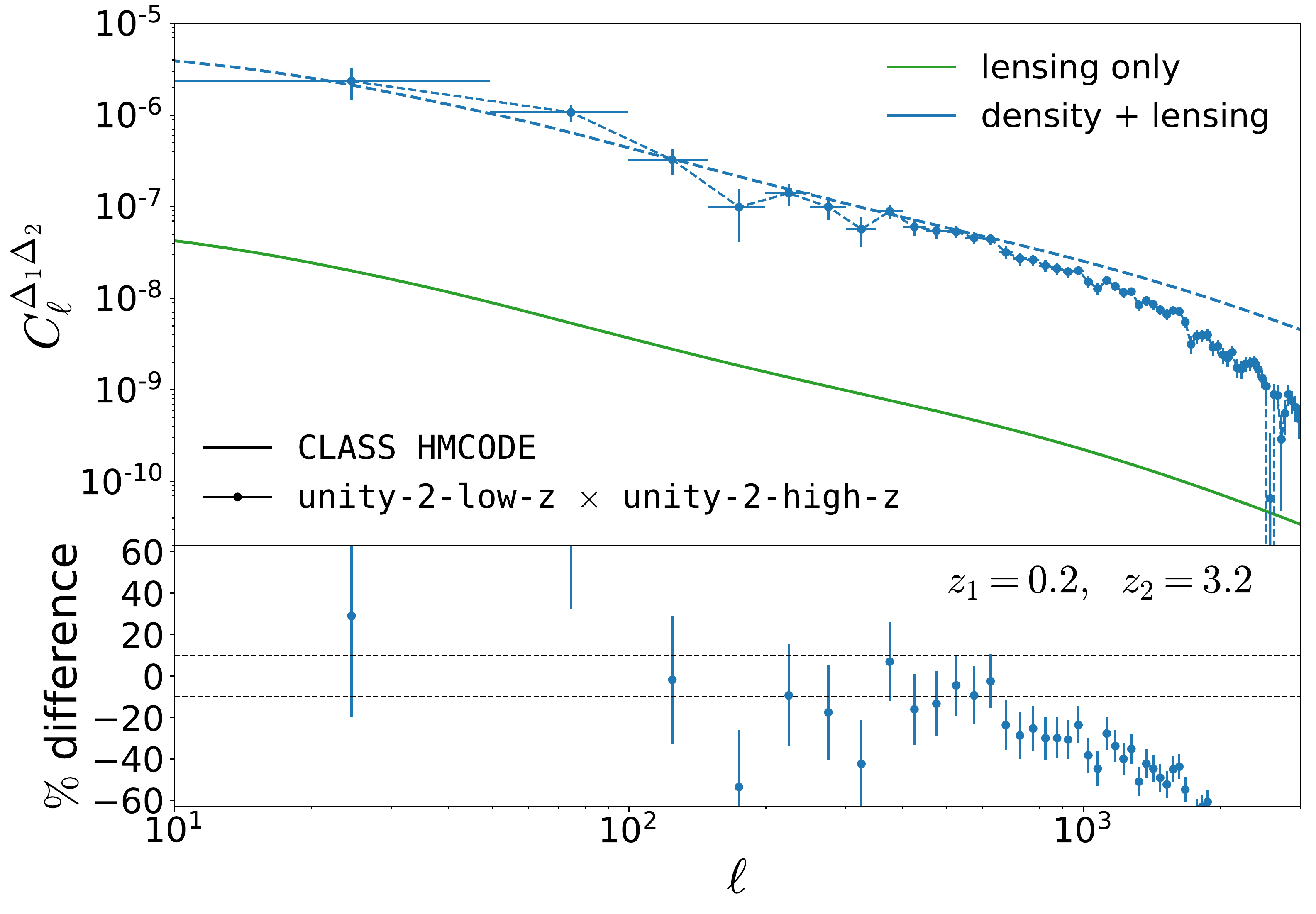}}
 \\[1ex]
   \parbox[c]{0.2\textwidth}{\subcaption{} \label{fig:lens-2} }\parbox[c]{0.8\textwidth}{\includegraphics[width=0.79\textwidth]{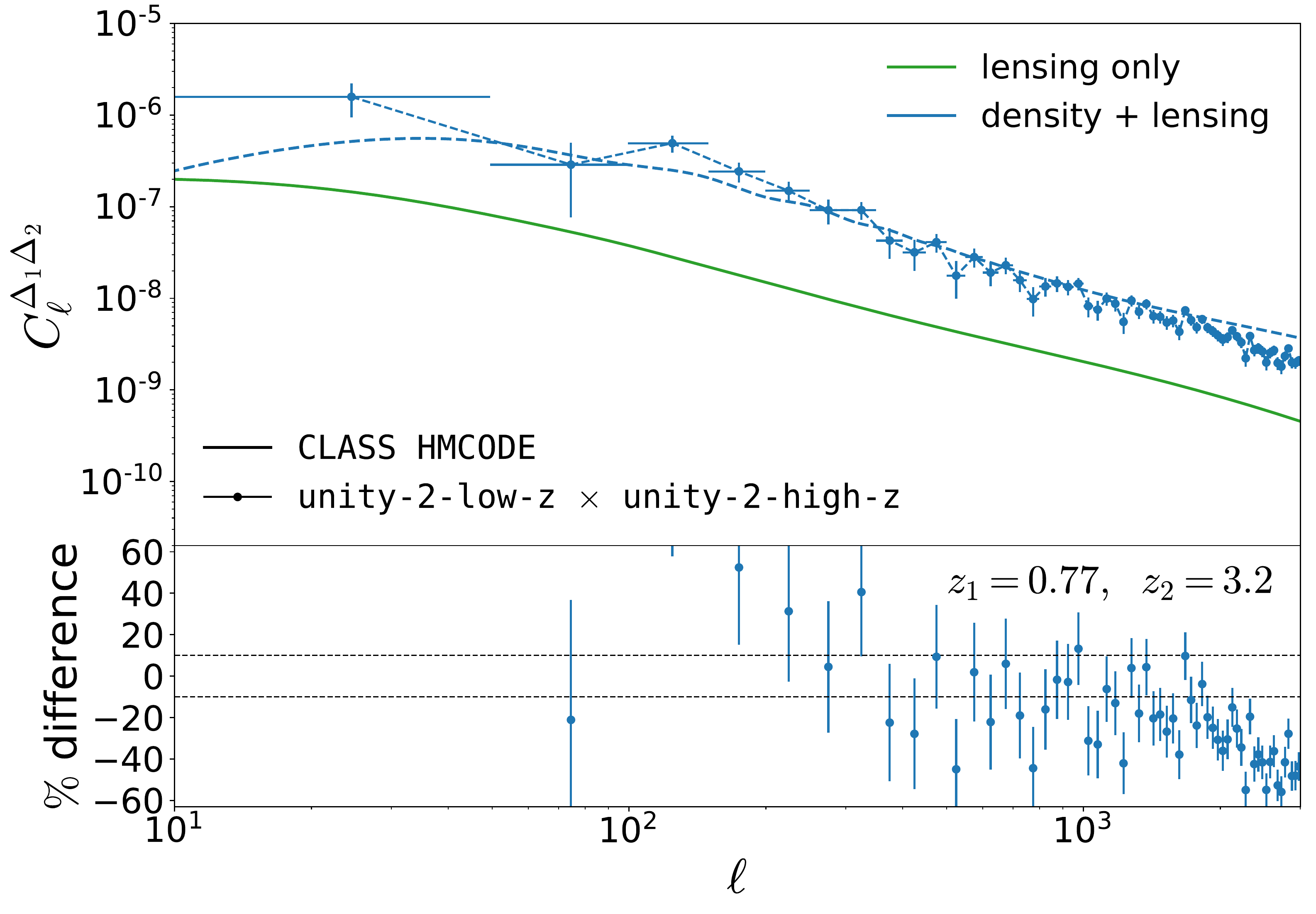}}
 \\[1ex]
          \caption{We show the unequal redshift correlations for two redshift pairs. The first case (a) with $z_1=0.2$ is dominated by the density-lensing correlations which are neglected in the green solid curve which shows the lensing-only contribution. For this lower redshift our simulations are missing power at $\ell>500$ due to finite resolution. For the second case (b) with $z_1=0.77$, the density-lensing signal is still about a factor 10 larger than the lensing-only term. For this redshift our simulations lose power for $\ell\gtrsim 1500$. The density-lensing signal is negative and indicated as dashed lines.}
         \label{fig:lens}    
\end{figure}

In order to study the non-linear regime of magnification, we consider the configuration in which this effect is by far the dominant contribution to the signal, i.e.\ the cross-correlations of bins with large redshift separation.
For this purpose we cross-correlate a map of the number counts from the \texttt{unity-2-high-z} catalogue at $z = 3.2$, with half-width $\sigma_\text{z} = 0.15$, with several maps from the 
\texttt{unity-2-low-z} catalogue. The low-redshift
maps were extracted from redshift bins
centered at $z = 0.2, 0.475, 0.6, 0.69, 0.77$
and with half-width $\sigma_\text{z} = 0.2, 0.075, 0.05, 0.04, 0.04$, respectively. 
The high-redshift map that we use for this analysis has a sky coverage $f_\text{sky} \sim 0.047$, while the low-redshift catalogue covers the full-sky. 
However, we apply a mask on the low-redshift catalogue, so that we only include pixels from half of the sky. 
This way we avoid spurious correlations at high separation angles, which would occur because the high-redshift survey footprint partially overlaps with the rear half of the low-redshift sky in the periodic domain of the \texttt{unity-2} simulation used here.

In Fig. \ref{fig:lens} we compare the results from 
our simulation with the {\sc class} predictions, from the cross-correlations of redshifts $z_1 = 0.2, z_2 = 3.2$ and  $z_1 = 0.77, z_2 = 3.2$.
The error-bars of the cross correlation signal are due to cosmic variance, and they have been estimated as
\begin{equation}
\sigma_{C^{\Delta_1 \Delta_2}_\ell} =
\sqrt{ \frac{C^{\Delta_1 \Delta_1}_\ell C^{\Delta_2 \Delta_2}_\ell + \left(C^{\Delta_1 \Delta_2}_\ell\right)^2}{f_\text{sky} \Delta \ell (2 \ell + 1)}},
\end{equation}
where $\Delta \ell$ is the width of the $\ell$-bins, while we use for $f_\text{sky}$ the sky coverage
of the high-redshift catalogue, i.e.\ we consider the
sky overlap of the two catalogues. These errors are dominated by the much larger auto-correlation term (the first term in the numerator).

In the plot, we do not show the contribution from the over-density of particles alone: it is several orders of magnitude smaller than the plotted cross-correlation and therefore negligible. 
The main contribution to the signal is the cross-correlation between the density at low redshift and lensing magnification of the high-redshift bin. The lensing-lensing contribution is negligible for small $z_1$, while we expect it to become more and more relevant as we cross-correlate bins at higher redshift.  Already for $z_1=0.77$, its contribution is at the level of 10\% and more.
The results from our simulation are affected by finite resolution at high $\ell$. 
The resolution of the simulation $\Delta x = 700\,\text{kpc/h}$ translates into a Nyquist wavenumber of
$k_\text{max} = \pi/\Delta x \approx 4.5\, h/\text{Mpc}$, which can be converted into the maximum multipoles that we can resolve $\ell_\text{max}\sim r(z) k_\text{max}$.
At $z = 0.2$ we have $\ell_\text{max} \sim 2560$. 
Therefore, in Fig.~\ref{fig:lens-1} we see the numerical oscillations at $\ell > 2500$ due to finite grid resolution of the simulation. As explained earlier, the error at $\ell_\text{max}(z)$ is of order unity and we expect a 5\% error already at about $\ell_{\max}/\sqrt{20}\simeq 570$.
In the configurations where lensing is important, finite-grid effects can contaminate the estimation of the angular power spectrum at high redshift due to the fact that lensing is an integrated effect and, therefore,  resolution error from smaller redshifts  propagate into the estimated magnification at all redshifts.
We have studied this in detail for the convergence field in Ref.~\cite{Lepori:2020ifz}, Appendix C. 
The  power suppression due to finite resolution on small scales, clearly visible in both panels of Fig.~\ref{fig:lens}, is caused by the finite resolution of our simulations. 

\begin{figure}[ht]
\captionsetup[subfigure]{labelfont={Large,bf}}
  \parbox[c]{0.2\textwidth}{\subcaption{}\label{fig:cross-1}}\parbox[c]{0.8\textwidth}{\includegraphics[width=0.79\textwidth]{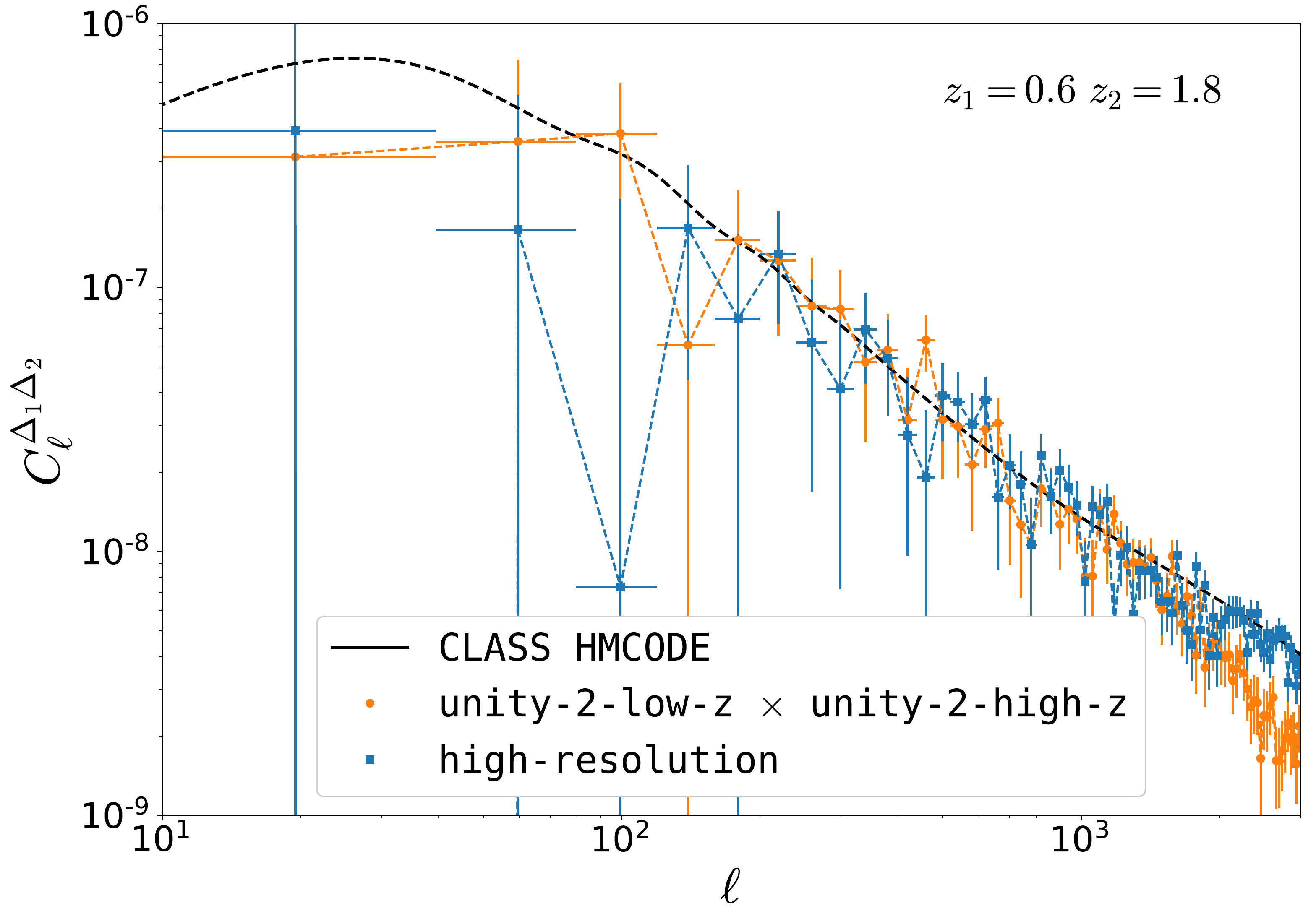}}
 \\[1ex]
   \parbox[c]{0.2\textwidth}{\subcaption{} \label{fig:cross-2} }\parbox[c]{0.8\textwidth}{\includegraphics[width=0.79\textwidth]{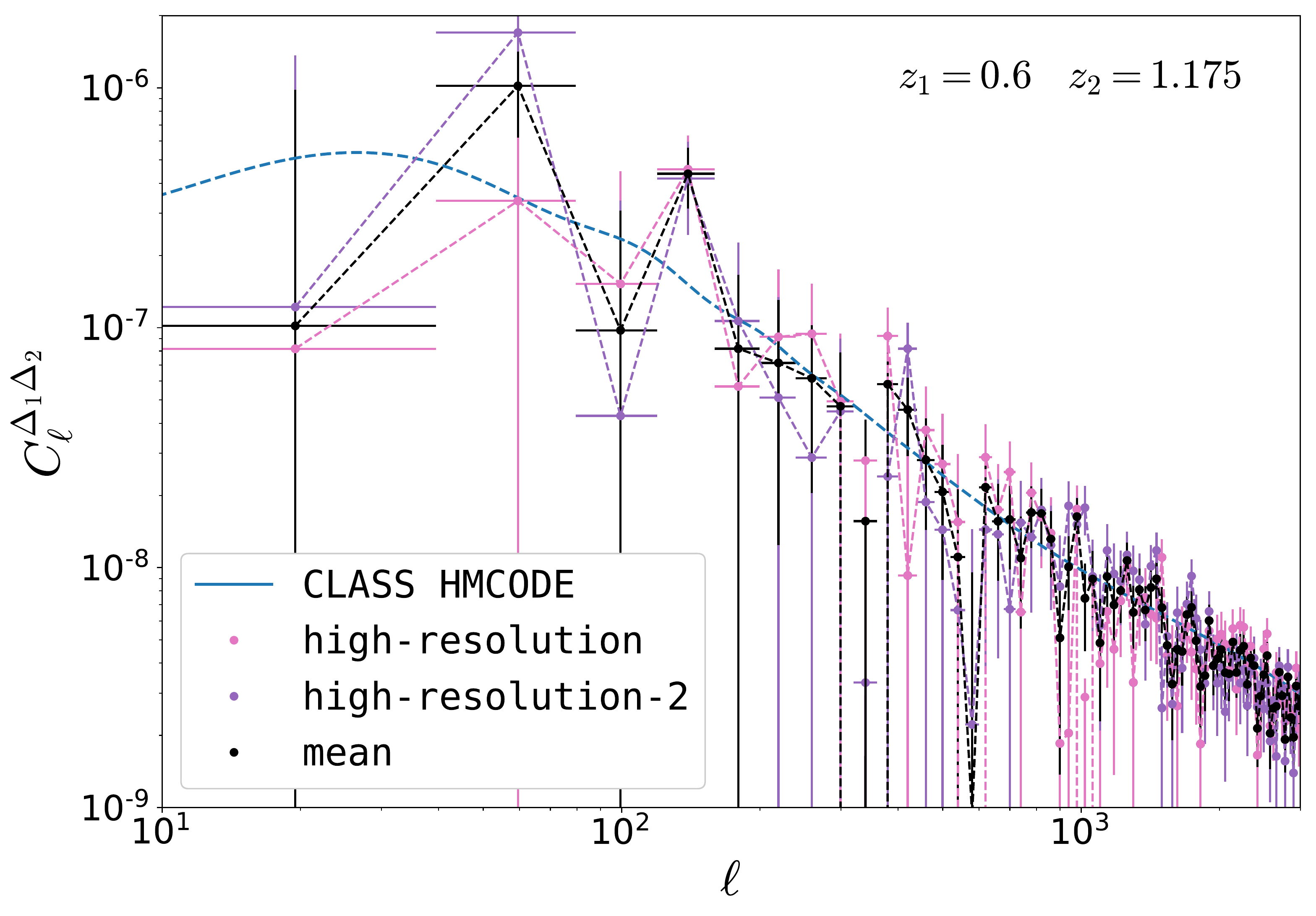}}
 \\[1ex]
          \caption{In the top panel we show the number-count cross-correlation spectrum for $z_1 = 0.6, z_2 = 1.8$ from the \texttt{high-resolution} catalogue and the \texttt{unity-2} catalogue. The latter has been computed cross-correlating the low-redshift catalogue and the high-redshift catalogues. In the bottom panel the cross-correlation for  redshifts $z_1 = 0.6, z_2 = 1.175$ are shown for the two high-resolution simulations. The black data points denote the mean of the two spectra. In both figures, the dashed lines indicate negative values of the theoretical power spectrum.
          }     \label{fig:cross}    
\end{figure}

\begin{table}
\begin{center}
\begin{tabular}{|c c c c| } 
 \hline
 $z$ & $\delta z$ & $N_\text{sources}$ (cat 1) & $N_\text{sources}$ (cat 2) \\ 
 \hline
 0.6 & 0.1 & 38342795  & 38068704\\
 0.8 & 0.1 & 54204035  & 53751424 \\
 0.95 & 0.05 & 31790875 & 31923809\\
 1.05 & 0.05 & 34578712 & 34982269 \\
 1.175 & 0.075 & 56375844  &56906758 \\
 1.325 & 0.075 & 61354137 & -\\
 1.475 & 0.075 & 65568421  &-\\
 1.625 & 0.075 & 67943255 &-\\ 
 1.8   & 0.1 & 93919982 & -\\ 
 \hline
\end{tabular}
\caption{\label{table:cross-binning}The mean redshifts and half-widths of 
the  bins considered in the cross-correlation analysis of the high-resolution simulations. We also quote the number of observed sources $N_\text{sources}$ in each redshift bin for the two catalogues.}
\end{center}
\end{table}

We have also extracted the number-count cross-correlation in our high-resolution simulations.  
The redshift binning for the two high-resolution catalogues are given in Table~\ref{table:cross-binning}.
The high-resolution catalogues have a sky coverage of $\sim 2.5\%$ of the sky. Therefore, when possible, we will show the angular power spectra estimated from two independent realisations. A comparison between them should give an indication of the impact of the large-scale statistical fluctuations. This source of uncertainty is investigated in detail in Appendix \ref{a:win}.
In Fig.~\ref{fig:cross-1} we show the cross-correlation for $z_1=0.6$ and $z_2=1.8$, for the \texttt{high-resolution}
run (blue) and for the \texttt{unity-2} run (orange). 
The dominant contribution to the angular power spectrum is the negative cross-correlation between density at low redshift and magnification at high redshift.
While the two spectra agree on intermediate scales, i.e.\ $150 < \ell < 1000$, on large scales the high-resolution 
spectrum exhibits a strong loss of power. This effect is not physical and it is due to the interplay of statistical fluctuations and finite sky coverage. 
On the other hand, on small scales, the spectrum estimated from the \texttt{unity-2} run shows a suppression for $\ell > 1000$, due to finite resolution. In this range of scales, the results from the \texttt{high-resolution} catalogue are more reliable and agree well with the theoretical prediction from {\sc class}.

In Fig.~\ref{fig:cross-2} we show the cross-correlation of two far-away redshift bins from the high-resolution catalogues. The black data points highlight the average between the spectra of the two independent realisations. 
Similarly to the configuration in Fig.~\ref{fig:cross-1}, the signal is dominated by the cross-correlation of density 
in the redshift bin $z_1$ with the lensing contribution at $z_2$. 
\subsection{Galaxy-galaxy lensing}

\label{sec:conv-cross}
\begin{figure}[!ht]
\captionsetup[subfigure]{labelfont={Large,bf}}
  \parbox[c]{0.2\textwidth}{\subcaption{}\label{fig:cross-kappa-1}}\parbox[c]{0.8\textwidth}{\includegraphics[width=0.79\textwidth]{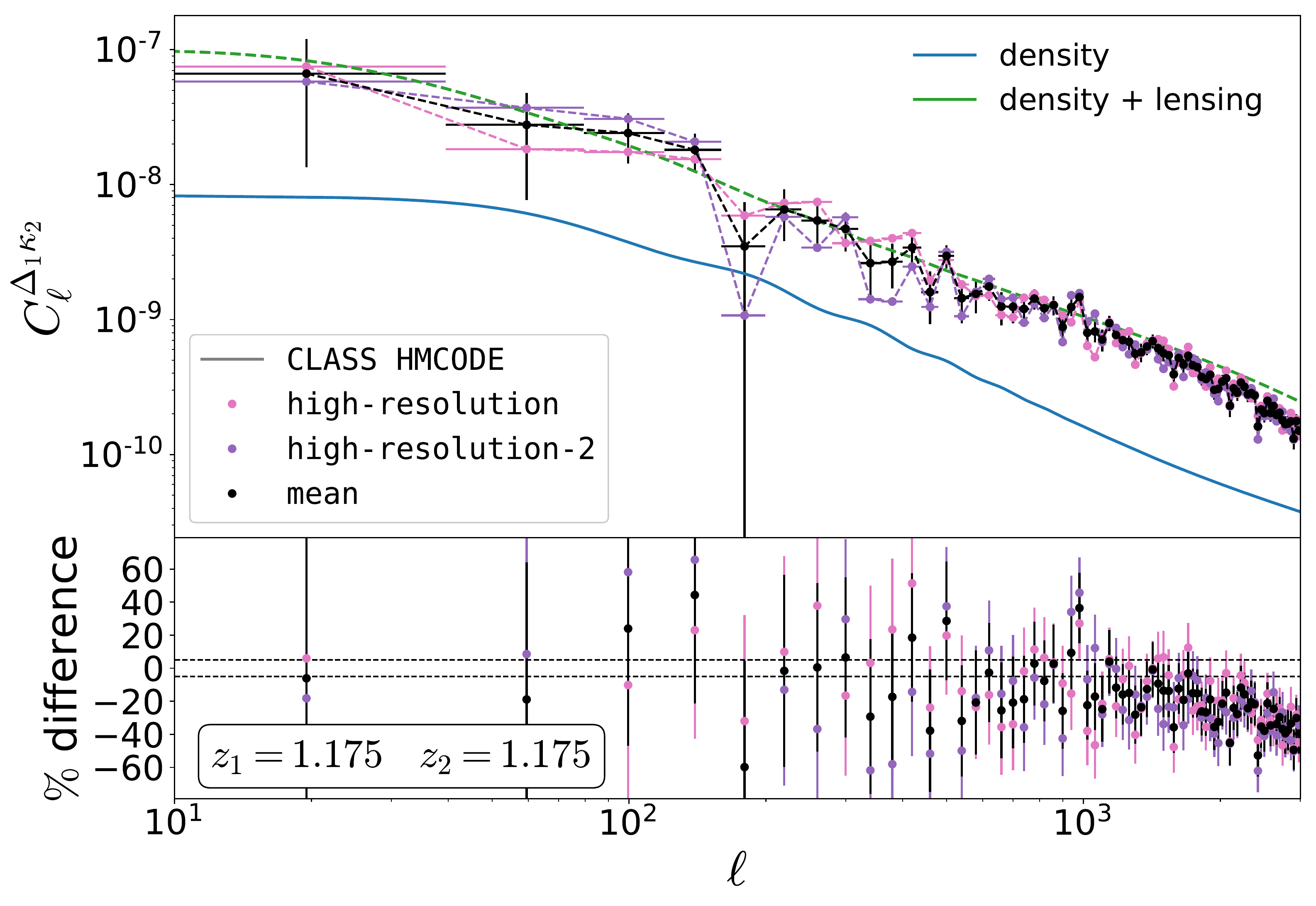}}
 \\[1ex]
   \parbox[c]{0.2\textwidth}{\subcaption{} \label{fig:cross-kappa-2} }\parbox[c]{0.8\textwidth}{\includegraphics[width=0.79\textwidth]{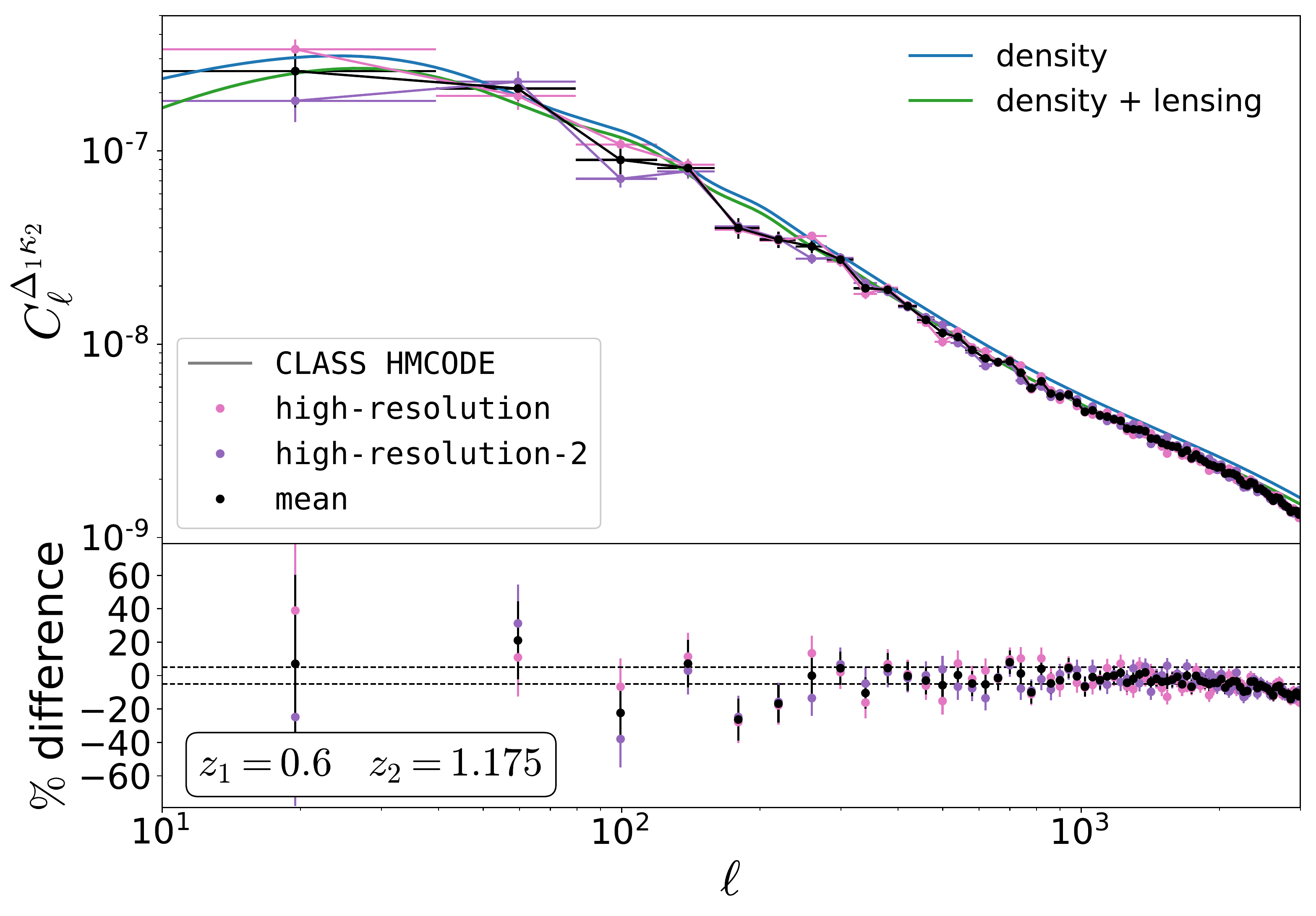}}
 \\[1ex]
\caption{ \label{fig:cross-kappa}    The $\Delta$-$\kappa$ cross-correlations are shown for two equal redshifts $z_1=z_2=1.175$ in the top panel and for two unequal redshifts, $z_1=0.6$ and $z_2=1.175$ in the bottom panel.}
\end{figure}

We have also investigated the cross-correlation spectrum of  number counts at redshift $z_1$ with the convergence field at redshift $z_2$, see Fig.~\ref{fig:cross-kappa}. This correlation can be measured by correlating shear measurements and number counts. It has been noted, see e.g.~\cite{Ghosh:2018nsm}, that this correlation does not only contain the density-$\ka$ term, but also a  $\kappa$-$\kappa$ (lensing-lensing) term. Note that this term is negative in $C_\ell^{\De\ka}$ since the number counts are proportional to $-2\kappa$. 
When we correlate a convergence map at $z_2$ with a number counts map at $z_1 > z_2$, this correlation spectrum is  mainly given by the lensing-lensing term because background overdensities are not correlated with the foreground lensing, while for $z_2>z_1$, both the density-lensing and the lensing-lensing contributions are relevant. 
A first attempt to measure the lensing-lensing cross correlation by correlating the foreground convergence field with background galaxies has recently been made with the Hyper Suprime-Cam Subaru Strategic Program
(HSC-SSP)  data~\cite{Liu:2021gbm}. 
In Fig.~\ref{fig:cross-kappa-1} we show the cross-spectrum for $z_1=z_2=1.175$ for the two independent high-resolution catalogues (pink and purple lines). 
The black data points denote the mean of the two measurements and its error is computed as  $\sigma_{C^{\Delta_1 \kappa_2}_\ell}/\sqrt{2}$, where $\sigma_{C^{\Delta_1 \kappa_2}_\ell}$ is the error on the cross-correlation for a single realisation, which has been estimated similarly to the error on the cross-correlation of the number counts at unequal-redshift bins. 
Blue and green lines denote theoretical predictions from {\sc class}: the blue line is the cross-correlation of the convergence map and the density contribution to the number counts, while the green line includes the cross-correlation of $\kappa$
with both density and magnification terms to the number counts. The full result is negative and dominated by the lensing-lensing term.  

In Fig.~\ref{fig:cross-kappa-2} we show the cross spectrum for $z_1=0.6$ and $z_2=1.175$, using the same colours and conventions as in Fig.~\ref{fig:cross-kappa-1}. 
This spectrum is dominated by the density-lensing cross correlation, which is positive in  $C_\ell^{\De\ka}$. On large scales, the negative lensing-lensing term leads to a small suppression of the result. Above $\ell\sim 1000$ our simulations lose power due to finite resolution: at $\ell = 3000$ the suppression is $\sim 50\%$ of the spectrum in the equal-redshift configuration, whereas it 
makes 20\% of the signal for the cross-correlations of unequal-redshift bins, where the amplitude of the power spectrum itself is larger.

\section{Conclusions}\label{s:con}
We have studied the number-count  angular power spectrum from relativistic N-body simulations. As our sensitivity to very large angular scales is limited, we have mainly seen the contributions which dominate on intermediate and small scales: the density, RSD and lensing terms. 
We have found that density and lensing are well modelled by the linear perturbative description currently implemented in the Boltzmann code {\sc class}, over a wide range of scales. On the other hand, the same approach does not work for RSD. 

For moderately slim redshift bins the N-body results deviate by up to 80\% from the naive Kaiser prescription implemented in {\sc class}. For more narrow redshift bins where RSD are more prominent, this discrepancy is expected to be even larger. We have also found that non-linearities from RSD are relevant at the few-percent level already on very large scales, confirming previous results obtained with Newtonian simulations \cite{Fosalba:2013wxa} and 
analytical models for non-linear RSD in harmonic space \cite{Gebhardt:2020imr}.
An accurate modelling of RSD is therefore very important for the analysis of large galaxy surveys.

The cross-correlations of different redshift bins allow an accurate measurement of the density-lensing correlation which is characteristic of the theory of gravity and will provide an excellent test of General Relativity. We have also correlated convergence maps with the number counts. For a number counts map in the foreground these are also dominated  by the  density-lensing contribution while for a number counts map in the background and even at equal redshift they are dominated by the lensing-lensing term. Comparing this power spectrum with the corresponding shear power spectrum will represent another interesting test of General Relativity.

Despite the fact that we employ simulations and ray tracing that are relativistic at every stage, and therefore never invoke a Newtonian limit, we expect that our findings would be replicated very accurately in a Newtonian setting. Possible gauge issues would only need to be considered on very large angular scales where a linear treatment can be applied. The correction to the number counts to adjust for the chart in a Newtonian simulation can be worked out from the treatment presented in \cite{Adamek:2019aad}, but we expect it to be smaller than cosmic variance in most practical circumstances. However, the situation can be much different in models beyond $\Lambda$CDM where the metric perturbations (and hence the lensing and RSD) may be strongly modified, e.g.\ by the presence of a dynamical dark energy field \cite{Hassani:2020buk}. In such cases the Newtonian limit has to be examined carefully, and using a relativistic framework may indeed be preferable.

As we have analysed catalogues drawn directly from N-body particles in this work, we have determined the non-linear angular-redshift power spectra of cold dark matter and baryons.
We could also have looked at biased tracers instead, e.g.\ halos. However, then the effects of non-linearities and bias would be entangled in the analysis. We therefore preferred to first look at unbiased tracers. The impact of non-linear bias (including evolution and magnification biases) in the fully relativistic number-count statistics deserves a dedicated study.

\section*{Acknowledgements}

We thank Farbod Hassani for sharing simulation data that was used during early stages of the project, and Pablo Fosalba for providing helpful comments on the first version of this manuscript. This work was supported by a grant from the Swiss National Supercomputing Centre (CSCS) under project ID s1035. We acknowledge financial support from the Swiss National Science Foundation.

{\small\paragraph{Carbon footprint} 
Simulations and post-processing consumed approximately $11880~\mathrm{kWh}$ of electrical energy from the Swiss power grid. Using a conversion factor of $0.149\,\mathrm{kg\,CO_2\,kWh}^{-1}$ (taken from \href{https://co2.myclimate.org/en}{myclimate.org}\footnote{Retrieved 19.\ February 2021}) this caused emissions of $1770\,\mathrm{kg\,CO_2}$. Emissions are being fully offset through the carbon offsetting initiative of the Institute for Computational Science at University of Zurich, partnering with the myclimate Foundation.}

\section*{Disclaimer}

This is an author-created, un-copyedited version of an article published in the Journal of
Cosmology and Astroparticle Physics (JCAP). IOP Publishing Ltd is not responsible for any
errors or omissions in this version of the manuscript or any version derived from it. The
Version of Record is available online at \url{https://doi.org/10.1088/1475-7516/2021/12/021}.

\appendix

\section{Effect of the window function}
\label{a:win}
Typically, observations (and also our simulations) do not cover the full sky, $4\pi$, but have only partial coverage. This is taken into account with a window function in the way which we now derive.

Given a two-point correlation function on the sphere, $\xi_{AB}(\mathbf{n}\cdot\mathbf{n}') = \left\langle \Delta_A(\mathbf{n}) \Delta_B(\mathbf{n}')\right\rangle$, its spherical harmonic coefficients $C^{AB}_\ell$ can be computed as

\begin{equation}
\label{eq:2pcftoCl}
    C^{AB}_\ell = \frac{1}{4 \pi} \iint d\Omega d\Omega' \xi_{AB}(\mathbf{n}\cdot\mathbf{n}') P_\ell(\mathbf{n}\cdot\mathbf{n}')\,,
\end{equation}
which can also be inverted to
\begin{equation}
\label{eq:Clto2pcf}
    \xi_{AB}(\mu) = \sum_\ell \frac{2\ell + 1}{4 \pi} C^{AB}_\ell P_\ell(\mu)\,,
\end{equation}
where $\mu = \mathbf{n}\cdot\mathbf{n}'$ is the scalar product of the two direction vectors.

If we have only partial sky coverage, we may introduce weights $w_A(\mathbf{n})$, $w_B(\mathbf{n}')$ that characterise the selection function, or mask, and write uncorrected harmonic coefficients as
\begin{equation}
\label{eq:uncorrectedCl}
    \tilde{C}^{AB}_\ell = \frac{1}{4 \pi} \iint d\Omega d\Omega' w_A(\mathbf{n}) w_B(\mathbf{n}') \xi_{AB}(\mathbf{n}\cdot\mathbf{n}') P_\ell(\mathbf{n}\cdot\mathbf{n}')\,,
\end{equation}
which can simply be estimated from the weighted data using the standard estimators developed for the full sky. Specifically, in the \texttt{PolSpice} code used in this work, the weighted map is expanded in spherical harmonic coefficients estimated from the pixel sum
\begin{equation}
    \tilde{\Delta}^{A,B}_{\ell m} \simeq \frac{4 \pi}{N_\mathrm{pixels}} \sum_{\mathbf{n} \in \mathrm{pixels}} Y^\ast_{\ell m} (\mathbf{n}) w_{A,B}(\mathbf{n}) \Delta_{A,B}(\mathbf{n})\,,
\end{equation}
which are then taken in a scalar product and averaged over $m$ to obtain the estimator of $\tilde{C}_\ell^{AB}$ as
\begin{equation}
    \tilde{C}_\ell^{AB} \simeq \frac{1}{2\ell + 1} \sum_m \tilde{\Delta}^A_{\ell m} \tilde{\Delta}^{B\ast}_{\ell m}\,.
\end{equation}
Note that this estimator coincides with Eq.~\eqref{eq:uncorrectedCl} only in the expectation value and in the limit of large number of pixels. We shall proceed with considering the expectation value for now but return to this point towards the end of this appendix. Applying Eq.~\eqref{eq:Clto2pcf} on the result also defines an uncorrected two-point correlation function $\tilde{\xi}_{AB}(\mu)$.

We may further define a power spectrum and corresponding correlation function of the mask as
\begin{equation}
    C^\mathrm{mask}_\ell = \frac{1}{4 \pi} \iint d\Omega d\Omega' w_A(\mathbf{n}) w_B(\mathbf{n}') P_\ell(\mathbf{n}\cdot\mathbf{n}')\,,
\end{equation}
which evidently has a deterministic estimator.

From these ingredients we can now construct a corrected and apodised correlation function $\hat{\xi}_{AB}(\mu)$ as
\begin{equation}
    \hat{\xi}_{AB}(\mu) = \frac{\tilde{\xi}_{AB}(\mu)}{\xi^\mathrm{mask}(\mu)} W(\mu)\,,
\end{equation}
where $W(\mu)$ is an apodisation function that needs to be zero in the range of $\mu$ where $\xi^\mathrm{mask}(\mu)$ is zero, i.e.\ where there are no pairs of pointings $\mathbf{n}$, $\mathbf{n}'$ with $\mu= \mathbf{n}\cdot\mathbf{n}'$ due to the mask. More generally the apodisation is applied to suppress the amplification of noise where $\xi^\mathrm{mask}(\mu)$ becomes small.

Finally, by applying again Eq.~\eqref{eq:2pcftoCl}, we arrive at the corrected power spectrum $\hat{C}_\ell^{AB}$ as estimated by the \texttt{PolSpice} code. In the following we would like to study the relation between $\hat{C}_\ell^{AB}$ and the ``true'' $C_\ell^{AB}$ of the full sky.

First, by inserting the Legendre expansion of Eq.~\eqref{eq:Clto2pcf} into Eq.~\eqref{eq:uncorrectedCl} we find that
\begin{equation}
    \hat{\xi}^{AB}(\mu) = \frac{W(\mu)}{\xi^\mathrm{mask}(\mu)}\sum_{n,m} P_n(\mu) \frac{2n+1}{(4\pi)^3} \left(2m+1\right) C_m^{AB} \!\iint d\Omega d\Omega' w_A(\mathbf{n}) w_B(\mathbf{n}') P_n(\mathbf{n}\cdot\mathbf{n}') P_m(\mathbf{n}\cdot\mathbf{n}')\,.
\end{equation}

Next, if we expand the weight functions into spherical harmonics, the double integral becomes the product of two Gaunt integrals. Specifically, if we write
\begin{equation}
   w_A(\mathbf{n}) = \sum_{L, M} w_{LM}^A Y_{LM} (\mathbf{n})\,, 
\end{equation}
and similarly for $w_B(\mathbf{n}')$, we find that
\begin{equation}
    \iint d\Omega d\Omega' w_A(\mathbf{n}) w_B(\mathbf{n}') P_n(\mathbf{n}\cdot\mathbf{n}') P_m(\mathbf{n}\cdot\mathbf{n}') =
    4 \pi\!\! \sum_{L = |n-m|}^{n+m} \sum_M w_{LM}^A (w_{LM}^B)^\ast
    \begin{pmatrix}
L & n & m\\
0 & 0 & 0
\end{pmatrix}^2\,.
\end{equation}
The harmonic coefficients $\hat{C}_\ell^{AB}$ can therefore be written as
\begin{eqnarray}
\hat{C}_\ell^{AB} &=& \sum_{n,m} \frac{2n + 1}{2} \!\int_{-1}^1 \!\frac{W(\mu) P_\ell(\mu) P_n(\mu)}{\xi^\mathrm{mask}(\mu)} d\mu \frac{2m+1}{4\pi} C_m^{AB} \!\! \sum_{L = |n-m|}^{n+m} \sum_M w_{LM}^A (w_{LM}^B)^\ast
    \begin{pmatrix}
L & n & m\\
0 & 0 & 0
\end{pmatrix}^2\,,\nonumber\\
&=& \sum_{n,m} K_{\ell n} Q_{nm} C^{AB}_m\,,
\label{eq:wind-cl}
\end{eqnarray}
where in the second line we define two kernel matrices $K_{\ell n}$, $Q_{nm}$ which depend only on the weights and on the apodisation,
\begin{eqnarray}
\label{eq:Kmatrix}
    K_{\ell n} &=& \frac{2n + 1}{2} \!\int_{-1}^1 \!\frac{W(\mu) P_\ell(\mu) P_n(\mu)}{\xi^\mathrm{mask}(\mu)} d\mu\,,\\
    Q_{nm} &=& \frac{2 m + 1}{4 \pi}\sum_{L = |n-m|}^{n+m} \sum_M w_{LM}^A (w_{LM}^B)^\ast
    \begin{pmatrix}
L & n & m\\
0 & 0 & 0
\end{pmatrix}^2\,.
\end{eqnarray}

The computation of the matrix elements of $K_{\ell n}$ can be simplified by defining a new function $\hat{W}(\mu) = W(\mu) / \xi^\mathrm{mask}(\mu)$, with harmonic coefficients $\hat{W}_{n'}$ computed in the usual way,
\begin{equation}
    \hat{W}_{n'} = 2 \pi \int^1_{-1} \frac{W(\mu)}{\xi^\mathrm{mask}(\mu)} P_{n'}(\mu) d\mu\,.
\end{equation}
With these harmonic coefficients, the integral in Eq.~\eqref{eq:Kmatrix} again becomes a Gaunt integral, and hence
\begin{equation}
    K_{\ell n} = \frac{2 n + 1}{4 \pi} \sum_{n'=|\ell-n|}^{\ell+n}\!\! \hat{W}_{n'} \left(2 n' + 1\right)
\begin{pmatrix}
\ell & n & n'\\
0 & 0 & 0
\end{pmatrix}^2\,.
\label{eq:kln}
\end{equation}

\begin{figure}[t]
\begin{center}
\includegraphics[width=0.75\textwidth]{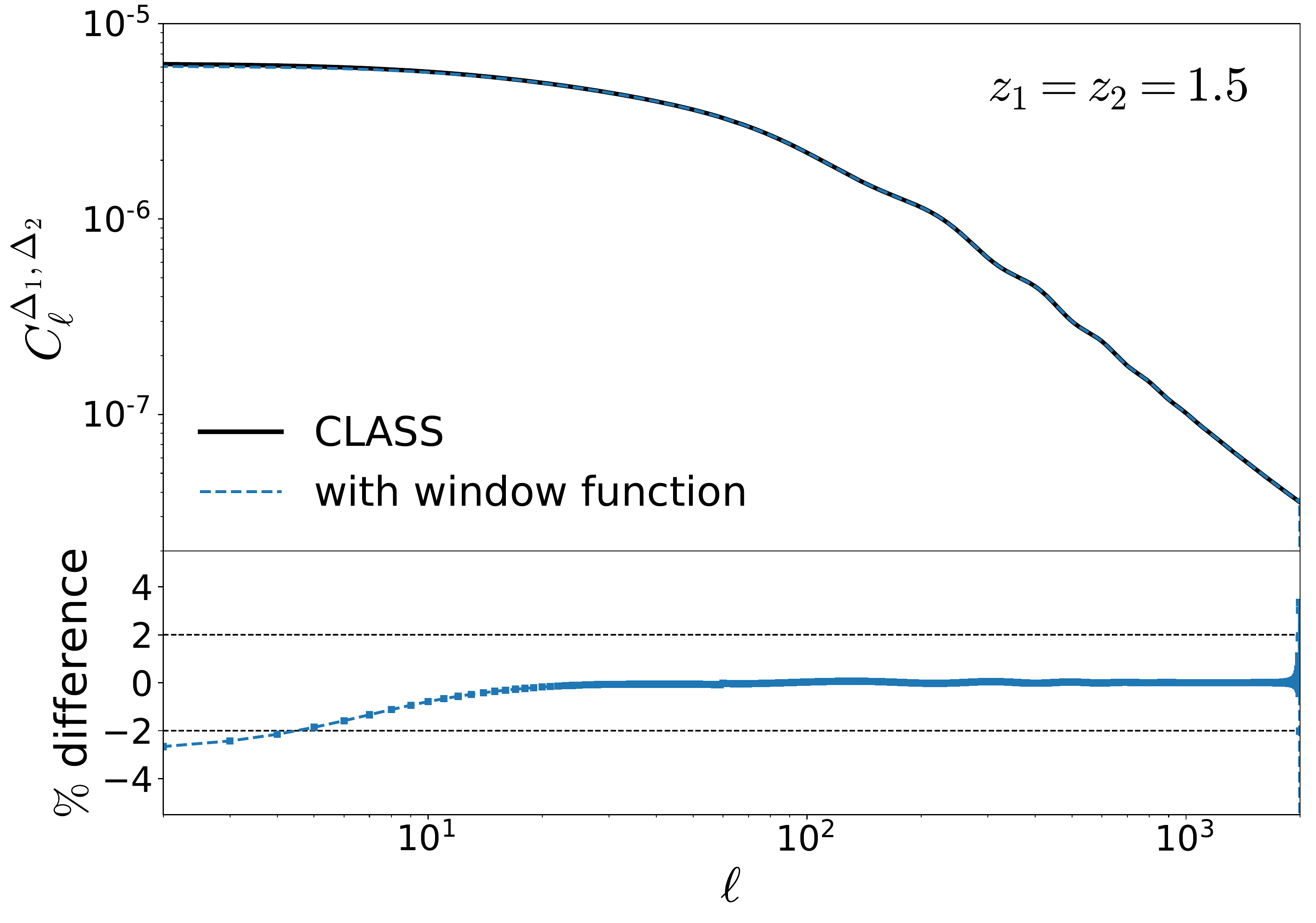}   
    \caption{\label{fig:window} Comparing the full sky {\sc class} result with a windowed spectrum with 2.5\% sky coverage calculated as described in the text. For $\ell>10$ the windowed result rapidly approaches the full sky spectrum.
    }
\end{center}
\end{figure}

In our case we have a particularly simple weight function $w_{A,B}(\mathbf{n}) = \theta(\mathbf{n}\cdot\mathbf{z} - \mu_o)$, where $\mathbf{z}$ is the unit vector pointing along the $z$-axis, and $\mu_o$ is the cosine of our opening half-angle. The spherical harmonic coefficients are
\begin{equation}
    w_{LM}^{A,B} = \delta_{M0} \sqrt{\frac{\pi}{2 L + 1}} \left[P_{L-1}(\mu_o) - P_{L+1}(\mu_o)\right]\,,
\end{equation}
where we use the standard definition that $P_{-1}(\mu) = 1$. We can also write a closed expression for $\xi^\mathrm{mask}(\mu)$,
\begin{equation}
    \xi^\mathrm{mask}(\mu) = \frac{1}{4}\sum_{\ell} P_\ell(\mu) \frac{\left[P_{\ell-1}(\mu_o) - P_{\ell+1}(\mu_o)\right]^2}{2\ell+1}\,.
\end{equation}
The kernel matrix $Q_{nm}$ becomes
\begin{equation}
    Q_{nm} = \frac{2 m + 1}{4}\!\! \sum_{L = |n-m|}^{n+m} \!\!\frac{\left[P_{L-1}(\mu_o) - P_{L+1}(\mu_o)\right]^2}{2L + 1}
    \begin{pmatrix}
L & n & m\\
0 & 0 & 0
\end{pmatrix}^2\,.
\label{eq:qnm}
\end{equation}

In Fig.~\ref{fig:window} we illustrate the effect of the window function on the angular power spectrum. The top panel shows the auto-correlation of the number counts at $z = 1.5$ that we would measure
from a full-sky catalogue (black continuous line) and the same power spectrum measured from a partially masked map. The latter is estimated from Eq.~\eqref{eq:wind-cl}, where the two kernels $K_{\ell n}$ and $Q_{nm}$ have been computed respectively from Eqs.~\eqref{eq:kln} and \eqref{eq:qnm} assuming an opening angle $\sigma = \pi/5$ and a cosine window in the form
\begin{equation*}
    W(\mu) = \frac{1+\cos{(\pi \alpha)}}{2}, \qquad \alpha = \frac{\arccos{\mu}}{\sigma}.
\end{equation*}
This setting corresponds to a sky coverage of $\sim 2.5\%$, which is the sky fraction covered by our \texttt{high-resolution} catalogues. 
In the bottom panel we show the difference between the two spectra: we find that the window function slightly suppresses the power spectrum estimator on large scales and that this effect is $\sim 2.5\%$ at $\ell < 10$, i.e. well within the cosmic variance uncertainty. 

\begin{figure}[ht]
\begin{center}
\includegraphics[width=0.75\textwidth]{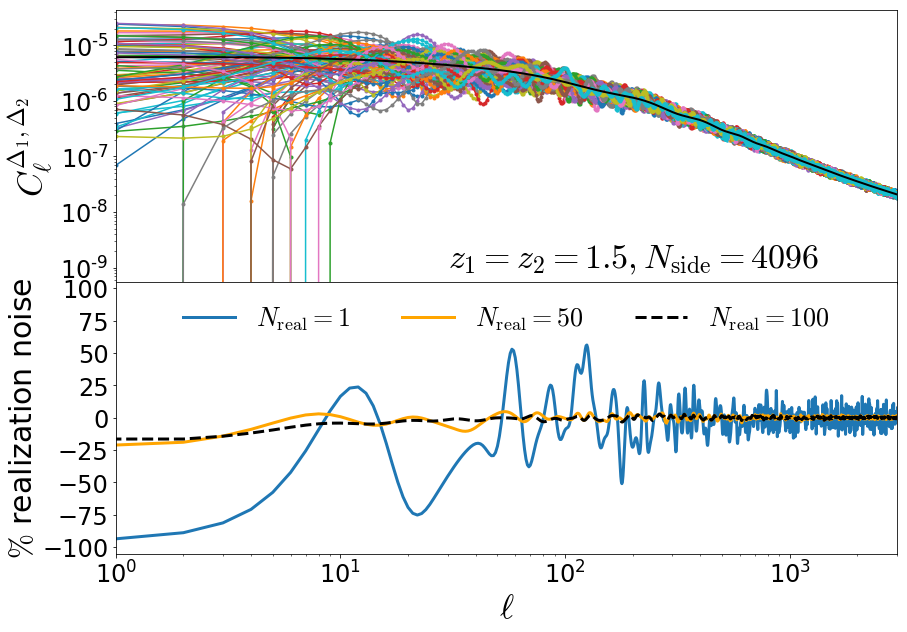}   
    \caption{\label{fig:real-noise} Statistical fluctuations of our estimator applied to a map with partial sky coverage ($f_\text{sky} = 0.025$). Top panel: true angular power spectrum (black) and angular power spectrum estimated from $100$ Gaussian realisations, after applying the mask (coloured lines). Bottom panel: realisation noise of our estimator for $N_\text{real}$ realisations.   
    }
\end{center}
\end{figure}

The windowed angular power spectrum estimator in Eq. \eqref{eq:wind-cl} represents the expectation value, i.e.\ the mean over an infinite number of realisations. 
However, in our observations and simulations we have a finite number of realisations. In order to quantify the \textit{realisation noise}, i.e.\ the amplitude of the fluctuations that we should expect when estimating the angular power spectrum from $N_\text{real}$ independent realisations, we test this effect with simulated maps. 
Starting from a theoretical spectrum at $z = 1.5$ 
we generate $100$ Gaussian realisations, we apply a Boolean mask in such a way that our maps cover only $~2.5\%$ of the sky, and we estimate for each realisation the angular power spectrum with \texttt{PolSpice}. In Fig.~\ref{fig:real-noise},
top panel, we show the full-sky theoretical spectrum in black, while the coloured lines are
the spectra estimated from the $100$ masked maps. 
On small scales, the amplitude of the fluctuations 
around the full-sky spectrum are small and average to zero, while on large scales the spectra estimated from different realisations can differ by orders of magnitude.
In the bottom panel we show the realisation noise for $N_\text{real}$ that we quantify as
\begin{equation}
\%\,\,\text{realisation}\,\,\text{noise} = 100 \times \frac{1}{C^\text{true}_\ell}  \sum_{j =1}^{N_\text{real}}   \frac{C^{j}_\ell - C^\text{true}_\ell}{N_\text{real}},
\end{equation}
where $C^\text{true}_\ell$ is the true power spectrum while the
$C^{j}_\ell$ are the power spectra estimated from the $100$ Gaussian realisations after applying our mask.
For a single realisation (blue line) large-scales fluctuations for a map with a sky coverage $f_\text{sky} = 0.025$ can be of order unity and are correlated significantly across a range of $\ell$. Considering many realisations, these statistical fluctuations are strongly suppressed: for $100$ realisation (dashed black line) they are $\sim 10-15 \%$
on $\ell < 10$ and negligible on much smaller scales. Summing over an infinite number of realisation, the realisation noise here estimated should average out and the result should converge to the effect of the window function plotted in Fig.~\ref{fig:window}.
Of course the realisation noise for $\ell\gtrsim10$ can also be reduced with a larger sky coverage.

\bibliographystyle{JHEP}
\bibliography{refs}

\end{document}